\documentclass[twocolumn, pra,showpacs,nofootinbib]{revtex4}
\usepackage{dcolumn,graphicx}
\begin{document}
\preprint{UNR Mar 2001-\today }
\title{ Fourth-order perturbative extension of the
 singles-doubles coupled-cluster method}
\author{Andrei Derevianko}
\email{andrei@unr.edu}
\author{ Erik D. Emmons}
\affiliation { Department of Physics, University of Nevada, Reno,
Nevada 89557}

\date{\today}

\begin{abstract}
Fourth-order many-body corrections to matrix elements for atoms with
one valence electron are derived. The obtained diagrams are
classified using coupled-cluster-inspired separation
into contributions from $n$-particle excitations from the lowest-order
wavefunction.
The complete set of fourth-order diagrams involves only
connected single, double, and triple excitations and disconnected
quadruple excitations.
Approximately half of the fourth-order
diagrams are {\em not} accounted for by the popular coupled-cluster method truncated
at single and double excitations (CCSD).
Explicit formulae are tabulated for the entire set of
fourth-order diagrams missed by the CCSD method and
its linearized version, i.e. contributions from connected triple
and disconnected quadruple excitations. A partial summation scheme
 of the derived fourth-order contributions to all orders of perturbation
theory is proposed.
\end{abstract}


\maketitle

\section{Introduction}
Atomic tests of the low-energy electroweak sector of the standard model
require both high-precision measurements and {\em ab initio} calculations of matching
accuracy.
%An advance in accuracy of {\em ab initio}
%calculations for univalent atoms is required due to
%an improved precision of measurements~\cite{WooBenCho97}
%of parity non-conserving (PNC) amplitude in $^{133}$Cs.
%High-precision {\em ab initio} atomic-structure calculations are vital, e.g., for
%atomic probes of ``new physics'' beyond the standard model of elementary particles.
%Experimental and theoretical uncertainties equally effect the
%errors in determination of nuclear weak charge.
The most precise measurement to date of parity violation in atoms
has been carried by Wieman and co-workers using $^{133}$Cs. The accuracy of this
experiment~\cite{WooBenCho97} is about 0.4\%, while  the relevant theoretical quantity
is calculated with 0.4-1\% uncertainty, depending on the authors's estimates~\cite{BenWie99,theorPNCallNew}.
A keen interest in
reducing the uncertainties is stimulated by a possible deviation
of the resulting nuclear weak charge from the prediction of the standard model.
This deviation was first reported in Ref.~\cite{BenWie99} and then
scrutinized in Refs.~\cite{theorPNCallNew}.
Very recent analyses~\cite{theorPNCallNew} of parity violation in $^{133}$Cs
focused on effects of the Breit interaction, vacuum polarization, and
neutron ``skin'', each contributing at the level of 0.2--0.6\%.
However, the effects of higher-order correlations  beyond
those considered in high-precision calculations by \citet{DzuFlaSus89} and \citet{BluJohSap90}
remain to be understood. Here we discuss in detail a possible extension to
the method employed in Ref.~\cite{BluJohSap90}.

The key to the 1\% accuracy achieved in Refs.~\cite{DzuFlaSus89,BluJohSap90}
lies in the application of all-order methods based on relativistic many-body perturbation
theory (MBPT).
These techniques, although summing certain classes of MBPT diagrams
to all orders of perturbation theory,
still  do not account for   an infinite number of residual diagrams.
It seems natural to augment a given all-order technique with some of the omitted
diagrams so that the formalism is complete through a certain order of MBPT.
To illustrate, the random-phase approximation (RPA)~\cite{AmuChe75}
fully recovers second order matrix elements but does not
subsume all third-order diagrams.
Among the omitted third-order contributions so called Brueckner-orbital diagrams
are known to be numerically as important as the RPA sequence
(see, e.g., discussions in Refs.~\cite{Sap98,JohLiuSap96}).

By the same virtue, certain diagrams starting from the
fourth order of MBPT are missed in
the popular coupled-cluster expansion~\cite{Ciz66,CoeKum60,LinMor86}
truncated at the single and
double level of excitations (CCSD), although  all third-order contributions
are recovered~\cite{BluJohLiu89}. It has been shown~\cite{BluJohSap91}
that one of the subsets of the fourth-order terms missed by the CCSD method
does contribute as much as a few per cent to Cs hyperfine-structure constants.
At the same time, the considered subset leads to  worse
theory-experiment agreement for electric-dipole amplitudes~\cite{SafJohDer99}.
We anticipate that
a systematic accounting of {\em all} omitted fourth-order contributions to matrix elements
in the CCSD method may lead to more accurate {\em ab initio} results.
Here we derive such complimentary fourth-order many-body contributions
for matrix elements.

%We focus on contributions to one-particle matrix elements
%(e.g. transition amplitudes or hyperfine structure constants)
%and investigate a relation of direct order-by-order
%MBPT with an all-order approach -- coupled-cluster method~\cite{Ciz66,CoeKum60,LinMor86}.

The paper is organized as follows. Basic starting formulas and notation of
many-body perturbation theory (MBPT) are
introduced in Section~\ref{Sec_basics}. The linked-diagram expansion
specialized to atoms with a single valence electron is discussed in
Section~\ref{Sec_LDE}. The derived wave-functions through the third order
of MBPT are discussed in Section~\ref{Sec_wf} and their relation
to the truncated coupled-cluster method in Section~\ref{Sec_CC}. Finally,
the derived fourth-order corrections to matrix elements are tabulated in the Appendix
and classification of the diagrams is given in Section~\ref{Sec_mel}.
Fig.~\ref{Fig_Z4} summarizes the results of our work.

The fourth-order expressions presented here
may be useful for an analysis of completeness of all-order methods
and for designs of a hierarchy of next-generation approximations
in atomic many-body calculations. As an example, we discuss
all-order generalizations of the derived fourth-order contributions.

\section{Partitioning of the atomic Hamiltonian}
\label{Sec_basics}
Here we briefly recap starting formulas of many-body perturbation
theory (MBPT) for atoms with one valence electron.
Our derivation in the fourth order of many-body perturbation
theory may be considered as an extension of the
work by \citet{BluGuoJoh87}. They presented formulas from
first-, second-, and third-order perturbation theory.
For the convenience of the reader we keep most of the original
notation from Ref.~\cite{BluGuoJoh87}.
% where the expressions up to the third
%order of MBPT were tabulated previously.

The many-body Hamiltonian of an atomic system may be represented as
\begin{eqnarray}
H = H_0 + V_{I} &=&
\left( \sum_{i}\,h_{\rm nuc}(\mathbf{r}_i) + \sum_{i}\, U_{\rm HF}(\mathbf{r}_i) \right)  + \nonumber \\
&+&\left( \frac{1}{2}\sum_{i \neq j}\frac{1}{r_{ij}}
-\sum_{i}\,U_{\rm HF}(\mathbf{r}_i) \right) \,,
\label{Eqn_Htot}
\end{eqnarray}
where $h_{\rm nuc}$ includes the kinetic energy of an electron and its
interaction with the nucleus, $U_{\rm HF}$
is the Hartree-Fock potential, and the last term represents the residual Coulomb interaction
between electrons. The summations go over all electrons in the system.
In MBPT the first part of the Hamiltonian is treated as the lowest-order Hamiltonian $H_0$
and the residual Coulomb interaction as a perturbation $V_I$.

For atoms with one valence electron outside a closed-shell
core the many-body wavefunction in the
lowest order $|\Psi_v^{(0)} \rangle$  is a Slater determinant
constructed from core and valence single-particle orbitals
$u_k$ which satisfy
\begin{equation}
\left( h_{\rm nuc}(\mathbf{r}) + U_{\rm HF}(\mathbf{r}) \right) u_k(\mathbf{r}) =
\varepsilon_k u_k(\mathbf{r}) \, .
\end{equation}
The solutions of the above one-particle equation form a basis for application
of the formalism of second quantization. In the second quantization
the lowest-order Hamiltonian $H_0$ and the perturbing residual Coulomb interaction $V_I$
may be expressed as
\begin{eqnarray}
H_0 &=& \sum_{i} \varepsilon_i a^\dagger_i a_i \, , \\
V_I &=& \frac{1}{2}\sum_{ijkl} g_{ijkl} a^\dagger_i a^\dagger_j a_l a_k
-   \sum_{ij} \left(U_{\rm HF} \right)_{ij} a^\dagger_i a_j  \, ,
\end{eqnarray}
where $a^\dagger_i$  and $a_i$  are creation and annihilation
operators for a one-particle state $i$.

The Coulomb integral $g_{ijkl}$ is conventionally defined as
\begin{equation}
g_{ijkl}=\int u_{i}^{\dagger}\left(  \mathbf{r}\right)  u_{j}^{\dagger}\left(
\mathbf{r}^{\prime}\right)  \frac{1}{\left|  \mathbf{r}-\mathbf{r}^{\prime
}\right|  }u_{k}\left(  \mathbf{r}\right)  u_{l}\left(  \mathbf{r}^{\prime
}\right)  d^{3}r\,d^{3}r^{\prime} \, .
\label{Eqn_CoulMel}
\end{equation}
The matrix elements of the Hartree-Fock potential may be expressed in terms
of the antisymmetrized Coulomb integral $\tilde{g}_{ijkl} =g_{ijkl} - g_{ijlk}$ as
\begin{equation}
\left(U_{\rm HF} \right)_{ij} = \sum_a \tilde{g}_{iaja} \, .
\label{Eqn_HFpot}
\end{equation}
Here the summation is over core orbitals; this potential is the so-called
frozen-core Hartree-Fock potential, i.e.,  first the core orbitals are
calculated employing the self-consistent Hartree-Fock procedure and then
the rest of the one-particle
states are obtained using Eq.~(\ref{Eqn_HFpot}) without varying the determined core orbitals.
Finally, in the language of second quantization the lowest-order wavefunction corresponds
to $|\Psi_v^{(0)}\rangle = a^\dagger_v | 0_c \rangle$, where $v$ labels the one-particle
state of the valence electron and the quasi-vacuum state $| 0_c \rangle$
describes the closed-shell core.

>From a practical standpoint derivation of MBPT expressions is greatly simplified
by the introduction of normal form of the operator products $N[ \ldots]$
and by a subsequent application of the Wick theorem~\cite{LinMor86}.
The notion of
normal products arises from separation of one-particle states  into two general
categories - occupied in the quasi-vacuum state $| 0_c \rangle$ ( i.e.,  core
orbitals enumerated by letters $a,b,c,d$ ) and complementary
excited states (indices $m,n,r,s$). Unspecified orbitals are labelled by indices
$i,j,k$, and $l$. In this scheme the one-particle
valence states $v$ and $w$ are classified as excited orbitals.

With the normal products
\begin{equation}
H_0 =  E_c^{(0)} + \sum_i \varepsilon_i \, N[a^\dagger_i a_i] \,
\end{equation}
 and
\begin{equation}
V_I = E_c^{(1)}+ \frac{1}{2}\sum_{ijkl} g_{ijkl} N[a^\dagger_i a^\dagger_j a_l a_k]\, ,
\label{Eqn_VInp}
\end{equation}
where $E_c^{(0)}= \sum_a \varepsilon_a$ and
$E_c^{(1)} = -\frac{1}{2} \sum_a \left(U_{\rm HF} \right)_{aa}$;
in the following discussion we omit these nonessential offset contributions.

It is worth noting that there is no one-body part of the perturbation $V_I$
present in Eq.~(\ref{Eqn_VInp}); this fact demonstrates the
utility of the frozen-core Hartree-Fock potential in  MBPT.
In Ref.~\cite{BluGuoJoh87},
the case of a model potential differing from $U_{\rm HF}$ was investigated explicitly
and it was found that the number of resulting diagrams is substantially larger than
in the Hartree-Fock case.
Due to the very large number of diagrams in the fourth order, here
we restrict our attention to the practically important frozen-core Hartree-Fock case.

\section{Linked-diagram expansion}
\label{Sec_LDE}
We proceed to the derivation of many-body contributions to wavefunctions using
the formalism of the generalized Bloch equation~\cite{LinMor86}. The Bloch
equation is formulated for the wave operator $\Omega_v$ which relates
the exact wavefunction
$|\Psi_v\rangle$ to the lowest-order result
 $|\Psi_v^{(0)}\rangle=a_{v}^{\dagger}|0_{c}\rangle$ as
\begin{equation}
|\Psi_v\rangle \, =\, \Omega_v\, |\Psi_v^{(0)}\rangle \, .
\end{equation}
It should be noted that as defined, this exact wavefunction is not normalized, rather
an intermediate normalization scheme $\langle\Psi_v^{(0)}|\Psi_v\rangle =1$ is employed
in the formalism. The exact correlation energy of the one-valence electron system is given by
\begin{equation}
\delta E = \langle\Psi_v^{(0)}|V_I \Omega_v|\Psi_v^{(0)}\rangle \, .
\label{Eqn_dE}
\end{equation}

The wave-operator satisfies the linked-diagram version of the generalized Bloch equation
\begin{equation}
\left[  \Omega_v,H_{0}\right]  =
\left\{  Q\, V_I \, \Omega_v - \left(\Omega_v-1\right)
P\, V_I \, \Omega_v \right\}_{\mathrm{linked}} \, ,
\label{Eqn_Bloch}%
\end{equation}
where the operator $P=|\Psi_v^{(0)}\rangle\langle\Psi_v^{(0)}|$ projects on the lowest-order
wavefunction and $Q=1-P$ is a complementary projection operator.
The subscript ``linked'' in the above equation prescribes that all the {\em unlinked}
Brueckner-Goldstone diagrams are to be discarded; a diagram is said to be unlinked
if it contains a disconnected part with no free lines other than valence lines. Finally,
$\left[  \Omega_v,H_{0}\right]$ is the commutator $\Omega_v\,H_{0} - H_{0}\,\Omega_v$.

The traditional Rayleigh-Schr\"{o}dinger perturbation theory is recovered
from the Bloch equation~(\ref{Eqn_Bloch}) by expanding the wave operator in
powers of the residual interaction $V_I$, $\Omega_v=\sum_{n=0} \, \Omega_v^{(n)}$.
The resulting recursive relation is~\cite{LinMor86}
\begin{eqnarray}
\left[  \Omega_v^{(n)},H_{0}\right]  &=&
\left\{  Q\, V_I \, \Omega_v^{(n-1)} -\right.  \nonumber\\
& & \left. \sum_{m=1}^{n-1} \Omega_v^{(n-m)}
P\, V_I \, \Omega_v^{(m-1)} \right\}_{\mathrm{linked}} \, .
\label{Eqn_RS_omega}
\end{eqnarray}
Here the iterations start with $\Omega_v^{(0)} = 1$. A
corresponding perturbative expansion of correlation energy reads
\begin{equation}
\delta E_v = \sum_{n=1} \delta E^{(n)}_v =
\sum_{n=1} \langle\Psi_v^{(0)}|V_I \Omega_v^{(n-1)} |\Psi_v^{(0)}\rangle \, .
\end{equation}

The last term on the r.h.s. of Eq.(\ref{Eqn_RS_omega}) gives rise to so-called ``folded''
or ``backward'' diagrams~\cite{LinMor86}. Instead of calculating the explicit
contributions of folded diagrams we use an all-order approach which incorporates their
effect in modified energy denominators. Such a reformulation allows for a direct
link to the coupled-cluster method outlined in Section~\ref{Sec_CC}.
The exact wave-operator $\Omega_v$ may be separated into valence and core parts,
$\Omega_v=\Omega_v^{\mathrm{val}}+\Omega^{\mathrm{core}}$, the $\Omega_v^{\mathrm{val}}$ part promoting a valence
electron from the $|\Psi_v^{(0)}\rangle$ determinant into an excited state.
$\Omega^{\mathrm{core}}$, describing excitations of core electrons, does not depend on any particular valence state.
Similarly, the correlation contribution to the total energy of the
system $\delta E_v$ may be broken into corrections to the
energies of valence and core electrons,
$\delta E_v=\delta E_v^{\mathrm{val}}+\delta E^{\mathrm{core}}$.

Suppose that the valence removal energy $\varepsilon_v+\delta E_v^{\mathrm{val}}$
is known at the desired order of perturbation theory (e.g., from coupled-cluster
calculations) or
from experiment. Projecting the original
Bloch equation~(\ref{Eqn_Bloch}) onto $|\Psi_v^{(0)}\rangle$
and using the definition of the projection operator $P$ together
with Eq.~(\ref{Eqn_dE}) for the correlation energy, one may show that
\begin{eqnarray*}
\lefteqn{\left[  \Omega_v,H_{0}\right]  |\Psi_v^{(0)}\rangle=}\\
&&\left\{  Q \, V_I \, \Omega_v\right\}
_{\mathrm{linked}}|\Psi_v^{(0)}\rangle-\delta E_v^{\mathrm{val}}\,\Omega_v
^{\mathrm{val}}|\Psi_v^{(0)}\rangle \, .
\end{eqnarray*}
Notice that the last term is represented by a product of two valence contributions,
since all other terms  produce unlinked diagrams.
Expanding the commutator and explicitly breaking the
term $\left\{Q\,V_I\Omega_v\right\}_{\mathrm{linked}}$ into
valence and core contributions we arrive at
\begin{eqnarray*}
\left(  \varepsilon_{v}+\delta E_v^{\mathrm{val}}-H_{0}\right)  \Omega_v
^{\mathrm{val}}|\Psi_v^{(0)}\rangle &  =\left(  \left\{Q \, V_I \, \Omega_v\right\}
_{\mathrm{linked}}\right)  ^{\mathrm{val}}|\Psi_v^{(0)}\rangle,\nonumber\\
\left(  \varepsilon_{v}-H_{0}\right)  \Omega^{\mathrm{core}}|\Psi_v^{(0)}\rangle
&  =\left(  \left\{Q \, V_I \, \Omega_v\right\}_{\mathrm{linked}}\right)
^{\mathrm{core}}|\Psi_v^{(0)}\rangle \, .
\label{Eqn_folded_split}
\end{eqnarray*}

Accounting for the folded diagrams in this way leads to an additional
shift $\delta E^{\mathrm{val}}$ in energy denominators of diagrams for the valence part of the
wave operator $\Omega_v^{\mathrm{val}}$. Mnemonically, every occurrence of the Hartree-Fock
energy of the valence electron $\varepsilon_{v}$ in the energy denominators
has to be replaced by the total removal energy
$\varepsilon_{v}+\delta E^{\mathrm{val}}$, since
$\left(  \varepsilon_{v}-H_{0}\right)  \Omega^{\mathrm{core}}|\Psi_v^{(0)}\rangle$ simplifies
to $\left(-H_{0}\right)  \Omega^{\mathrm{core}}|0_c\rangle$.
Keeping this rule in mind, we may combine the above equations
\begin{eqnarray}
\lefteqn{\left(  \varepsilon_{v}+\left(  \delta E_v^{\mathrm{val}}\right)  -H_{0}\right)
\Omega_v|\Psi_v^{(0)}\rangle= }\nonumber \\
&&\left\{  Q\, V_I \,\Omega_v\right\}_{\mathrm{linked}}|\Psi_v^{(0)}\rangle \, ,
\label{Eqn_Bloch_spec}
\end{eqnarray}
where $\left(  \delta E_v^{\mathrm{val}}\right)$ means that the $\delta E_v^{\mathrm{val}}$
correction should be included for the valence diagrams of $\Omega_v$ and
discarded otherwise.

We expand the wave operator in powers of the residual electron-electron
interaction $V_I$, $\Omega_v=\sum_{n=0}\Omega_v^{(n)}$ and obtain
\begin{eqnarray*}
(  \varepsilon_{v}+ (  \delta E_v^{\mathrm{val}} )  &-& H_{0} )
\Omega_v^{(n+1)}|\Psi_v^{(0)}\rangle =  \nonumber \\
&&\left\{  Q\, V_I \,\Omega_v^{(n)}\right\}_{\mathrm{linked}}|\Psi_v^{(0)}\rangle \, ,
\end{eqnarray*}
with $\Omega_v^{(0)}=1$. This equation may be interpreted as a linked-diagram
version of the Brillouin-Wigner perturbation theory for atoms with one valence
electron outside a closed core. Introducing the resolvent operator
\begin{equation}
R_{v}=\left(  H_{0}-\left[  \varepsilon_{v}+\left(  \delta E^{\mathrm{val}%
}\right)  \right]  \right)  ^{-1} \, ,
\end{equation}
we obtain (with $ |\Psi_v^{(n)} \rangle \equiv \Omega_v^{(n)} |\Psi_v^{(0)} \rangle$)
\begin{eqnarray*}
|\Psi_v^{(n)}\rangle&=&-R_{v}\left\{  Q\,V_I\,|\Psi_v^{(n-1)}\rangle\right\}
_{\mathrm{linked}}= \nonumber \\
& &\left(  -1\right)  ^{n}\left(  \left\{  R_{v}\,Q\, V_I\,\right\}
_{\mathrm{linked}}\right)  ^{n}|\Psi_v^{(0)}\rangle \, .
\end{eqnarray*}
>From this recursion relation we may generate corrections to wave functions at any
given order of perturbation theory. In practice, the derivation is rather tedious
and error-prone. We employed the symbolic algebra system Mathematica~\cite{Wol99} to derive
the expressions presented in this work.

\section{Wavefunctions through the third order of MBPT}
\label{Sec_wf}
For the derivation of fourth-order matrix elements one requires contributions
to wavefunctions through the third order. Expressions for
$|\Psi_v^{(n)}\rangle= \Omega_v^{(n)} |\Psi_v^{(0)}\rangle$ through the second order may be found in Ref.~\cite{BluGuoJoh87}.
Although we fully derived $|\Psi_v^{(3)}\rangle$, to keep the
manuscript to a manageable size, we present below only a qualitative discussion of
the third-order correction to the wave-function.

\begin{figure}[h]
\begin{center}
\includegraphics*[scale=0.75]{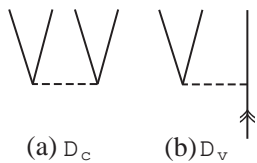}
\caption{Brueckner-Goldstone diagrams for the first-order
wave operator $\Omega_v^{(1)}$. Horizontal dashed lines represent residual Coulomb
interaction between electrons and vertical lines are particle/hole lines. The valence
line is marked by double arrow.\label{Fig_Om1}}
\end{center}
\end{figure}

The contributions to the wave operator $\Omega_v$ are conventionally classified
by the number of excitations from a reference determinant
$|\Psi_v^{(0)}\rangle = a^\dagger_v |0_c \rangle$. The  first-order
result, $\Omega_v^{(1)}$,
contains only double excitations drawn in Fig.~\ref{Fig_Om1}.
We may distinguish between valence and core excitations. The former
promote the valence electron to
an excited state ($\Omega_v^\mathrm{val}$) and the latter do not modify the
state of valence electron ($\Omega^\mathrm{core}$). With such a classification the
diagram Fig.~\ref{Fig_Om1}(a) represents core doubles $D_c$
and Fig.~\ref{Fig_Om1}(b) valence doubles $D_v$.

\begin{figure}[h]
\begin{center}
\includegraphics*[scale=0.75]{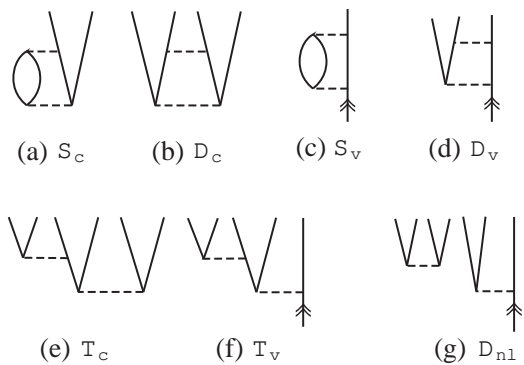}
\caption{ Sample contributions to the second-order
wave operator $\Omega_v^{(2)}$. \label{Fig_Om2}}
\end{center}
\end{figure}

\begin{figure}[h]
\begin{center}
\includegraphics*[scale=0.75]{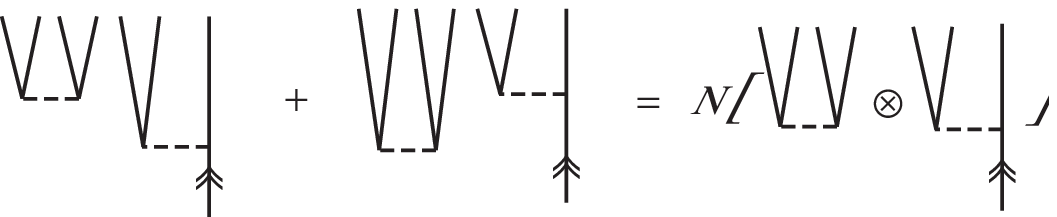}
\caption{
A sum of the the quadruple contribution Fig.~\protect\ref{Fig_Om2}(g) and a
similar diagram with the
order of two interactions reversed factorizes into a normal product of
double excitations. On the r.h.s. the energy denominators are to
be evaluated separately.
 \label{Fig_Dnl}}
\end{center}
\end{figure}

The second-order operator $\Omega_v^{(2)}$
contains  excitations up to quadruples.
%Complete set of diagrams of
%wave-operator $\Omega$ in the first and second order could be found in
%Ref.~\cite{LinMor86}.
Examples of contributions to $\Omega_v^{(2)}$ are drawn in Fig.~\ref{Fig_Om2}.
Diagrams~\ref{Fig_Om2}(a) and (b) represent some of the second-order core
singles and doubles. Valence singles and doubles are drawn in Fig. \ref{Fig_Om2}(c) and (d)
respectively. Diagrams \ref{Fig_Om2}(e) and (f) represent
core and valence triple excitations, and (g) --- disconnected quadruple excitations.
A sum of the the quadruple contribution \ref{Fig_Om2}(g) and a similar diagram with the
order of the two interactions reversed is known to factorize into a normal product of
double excitations~\cite{LinMor86}; this is demonstrated in Fig.~\ref{Fig_Dnl}.
We classify the disconnected quadruple contribution~\ref{Fig_Om2}(g) as a nonlinear
contribution of double excitations to wavefunctions.

\begin{figure}[h]
\begin{center}
\includegraphics*[scale=0.65]{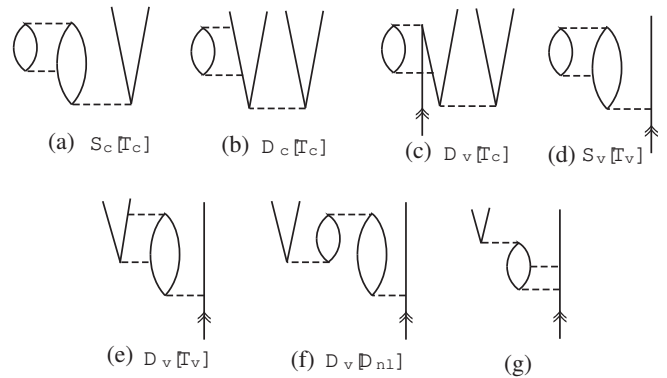}
\caption{ Representative diagrams for the third-order
wave operator $\Omega_v^{(3)}$.  \label{Fig_Om3}}
\end{center}
\end{figure}

Several contributions to the third-order wave-operator $\Omega_v^{(3)}$ are shown in
Fig.~\ref{Fig_Om3}. Single and double excitations shown in
Fig.~\ref{Fig_Om3}(a--e) contain intermediate triple excitations.
Diagram~\ref{Fig_Om3}(f) is due to intermediate second-order quadruple excitation.
For the sake of comparison with the coupled-cluster method we classify
diagram~\ref{Fig_Om3}(a) as the effect of core triples on core singles ($S_c[T_c]$),
(b) as modification of core doubles by core triples ($D_c[T_c]$), (c) as the
effect of core triples on valence doubles ($D_v[T_c]$), and (d) and (e) as the effect of
valence triples on valence singles and doubles ($S_v[T_v]$ and $D_v[T_v]$).
Finally, diagram~\ref{Fig_Om3}(f) may be classified as an effect of nonlinear doubles,
Fig.~\ref{Fig_Om2}(g), on valence doubles ($D_v[D_{nl}]$). It is worth noting that
the third-order wavefunction contains connected quadruple excitations and some
additional disconnected excitations; these corrections do not contribute
to the fourth-order matrix elements.

\section{Coupled-cluster method}
\label{Sec_CC}
The coupled-cluster (CC)  formalism~\cite{CoeKum60,Ciz66} is widely employed in atomic
and nuclear physics, and quantum chemistry~\cite{BisKum87}.
The main goal of the present work
is to identify fourth-order contributions to matrix elements not included in the truncated
singles-doubles
coupled-cluster method, and here we review the relevant features of this
all-order approach.
% The  {\em relativistic}
%atomic-structure CC-type calculations were performed for example in
%Refs.~\cite{BluJohLiu89,BluJohSap91,EliKalIsh94,AvgBec98,SafDerJoh98,SafJohDer99}.

The key point of the coupled-cluster method is the introduction of
an exponential ansatz for the wave operator~\cite{LinMor86}
\begin{equation}
 \Omega = N[ \exp(S)] = 1 + S + \frac{1}{2!} N[S^2] + \ldots \, .
 \label{Eqn_CCwo}
\end{equation}
The cluster operator $S=\Omega_\mathrm{conn}$ is expressed in terms of {\em connected} diagrams of the wave
operator $\Omega$, an example of disconnected
diagram being Fig.~\ref{Fig_Om2}(g). The operator $S$ is naturally broken
into cluster operators $S_n$ combining  $n$ simultaneous excitations
from the reference state $|\Psi_v^{(0)}\rangle$ in all orders of perturbation
theory.

Let us specialize the general formalism of Ref.~\cite{LinMor86} to the case of atoms
with one valence electron. A set of coupled equations for
the cluster operators may be found by considering connected diagrams
on both sides of the modified Bloch equation~(\ref{Eqn_Bloch_spec})
\begin{equation}
\left(  \varepsilon_{v}+\left(  \delta E_v^{\mathrm{val}}\right)  -H_{0}\right)
S_n =
\left\{  Q\, V_I \,\Omega_v\right\}_{\mathrm{conn, n}}\, ,
\end{equation}
where $\delta E_v$ is determined by Eq.~(\ref{Eqn_dE}) and wave operator
$\Omega_v$ by Eq.~(\ref{Eqn_CCwo}).
Term  $\left(  \delta E_v^{\mathrm{val}}\right)$ accounts for folded diagrams;
it is to be omitted for core and included for valence clusters.
Successive iterations of such all-order equations explicitly recover
order-by-order MBPT contributions to the wave operator discussed in the previous sections.

In most applications the full operator
$S$ is truncated at single and double excitations (CCSD method).
For univalent atoms the CCSD
parameterization may be represented as
\begin{eqnarray}
 S^\mathrm{SD}&=& S_1 + S_2 = \nonumber \\
 & &\sum_{ma} \rho_{ma}    \,  a^\dagger_m a_a +
 \frac{1}{2}  \sum_{mnab} \rho_{mnab} \, a^\dagger_m a^\dagger_n a_b a_a + \nonumber\\
 & &      \sum_{m \ne v} \rho_{mv}   \,   a^\dagger_m a_v +
 \sum_{mna} \rho_{mnva}    \,  a^\dagger_m a^\dagger_n a_a a_v \, ,
\end{eqnarray}
where the first two terms represent single and double excitations of core electrons
and the remaining contributions are valence singles and doubles.

It is worth emphasizing that the CCSD method is an all-order method. For example,
first--, second-- and third--order diagrams Fig.~\ref{Fig_Om1}(b),
Fig.~\ref{Fig_Om2}(d), and Fig.~\ref{Fig_Om3}(g)
are encapsulated in the valence doubles term
$\sum_{mna} \rho_{mnva}    \,  a^\dagger_m a^\dagger_n a_a a_v$.
Similarly, the CCSD method accounts for all single and double excitations (both
core and valence) shown in Figs.~\ref{Fig_Om1} and \ref{Fig_Om2}.
At the same time connected triple and higher-rank excitations
are not accounted for by the CCSD method, examples being Fig.~\ref{Fig_Om2}(e),(f)
and Fig.~\ref{Fig_Om3} (a--e). Although diagrams Fig.~\ref{Fig_Om3} (a--e)
are nominally single or double excitations, they contain connected
triples as intermediate excitations and are not included in the sequence
of CCSD diagrams.

A {\em linearized} version of the CCSD method (LCCSD) is a further simplification
of a hierarchy of all-order methods based on the coupled-cluster formalism.
In this approximation $\Omega_v^\mathrm{LCCSD} \equiv 1 + S^\mathrm{SD}$.
For alkali-metal atoms the LCCSD method was employed in
Refs.~\cite{BluJohLiu89,BluJohSap91,SafDerJoh98,SafJohDer99}.
Compared to the full CCSD approximation the linearized version misses a subset of
diagrams shown in Fig.~\ref{Fig_Om2}(g) and~\ref{Fig_Om3}(f).

To reiterate, connected triple excitations and disconnected
quadruple excitations first appear in the second order wavefunctions.
In order to systematically extend the CCSD method one has to investigate
the contributions of connected triple excitations and the role of nonlinear contributions
for the linearized CCSD approximation.

\section{Matrix elements}
\label{Sec_mel}
We investigate the fourth-order corrections to matrix element
of a one particle operator $Z=\sum_i z(\mathbf{r}_i)$. In second quantization
\begin{equation}
Z= \sum_{ij} z_{ij} a_i^\dagger a_j =  \sum_{a} z_{aa} +
\sum_{ij} z_{ij} N[a_i^\dagger a_j] \, ,
\end{equation}
where  $N[\ldots]$ denotes normal form of operator products.
We are mainly interested in matrix elements of non-scalar operators,
like electromagnetic transition amplitudes or pseudo-scalar operators,
like the electroweak interaction. For such operators the
contribution from the zero-body term $\sum_{a} z_{aa}$ vanishes and we
disregard it in the following discussion.

The exact matrix element between two valence states $w$ and $v$ is given
by
\begin{equation}
 M_{wv} =
\frac{ Z_{wv} }{\sqrt{ N_v N_w}} =
  \frac{
 \langle 0_c| a_w\Omega_w^\dagger \, Z \, \Omega_v a^\dagger_v|0_c\rangle
 }{\sqrt{ N_v N_w}} \, ,
 \label{Eqn_Zfull}
\end{equation}
where $\Omega_w$ and $\Omega_v$ correspond to wave operators for valence
states $w$ and $v$ respectively. Since the wave-operators were derived
using the intermediate normalization scheme, we introduced normalization factors
\[
N_v = \langle \Psi_v^{(0)} | \Omega_v^\dagger \Omega_v | \Psi_v^{(0)} \rangle
\]
in the definition of matrix element.

\citet{BluJohLiu89} have demonstrated that
disconnected diagrams in the perturbative expansion of the numerator and the denominator of
Eq.~(\ref{Eqn_Zfull}) cancel. Their final expression for the exact
matrix element reads
\begin{eqnarray}
M_{wv}&=&\delta_{wv}\left(  Z^{\mathrm{core}}\right)_{\mathrm{conn}}+ \nonumber \\
& & \frac{\left(  Z_{wv}^\mathrm{val} \right)_{\mathrm{conn}}
}{\left\{  \left[  1+\left(  N_{v}^\mathrm{val} \right)_{\mathrm{conn}
}\right]  \left[  1+\left(  N_{w}^\mathrm{val} \right)_{\mathrm{conn}
}\right]  \right\}  ^{1/2}} \, ,
\label{Eqn_Zconn}
\end{eqnarray}
where
%\begin{equation}
\[
Z^{\mathrm{core}} \equiv
\langle 0_c| \Omega_w^\dagger \, Z \, \Omega_v |0_c\rangle =
\langle 0_c| \left(\Omega^\mathrm{core}\right)^\dagger \, Z \,
\Omega^\mathrm{core} |0_c\rangle \,
\]
%\end{equation}
and the remaining contributions of
$Z_{wv}=\langle 0_c| a_w\Omega_w^\dagger \, Z \, \Omega_v a^\dagger_v|0_c\rangle$
are grouped into the valence part $ Z_{wv}^\mathrm{val}$. The diagrams of
$ Z_{wv}^\mathrm{val}$
explicitly depend on valence indices $w$ and $v$. The valence part of the normalization
factor $N_{v}^\mathrm{val}$ is defined in a similar fashion.
%Since
%the total angular momentum of a closed-shell core is zero,
The core contribution $Z^{\mathrm{core}}$ vanishes for non-scalar (and pseudo-scalar)
operators and we disregard $Z^{\mathrm{core}}$ in the
following discussion.
Notice that all the diagrams in Eq.~(\ref{Eqn_Zconn})
must be rigorously connected as emphasized by  subscripts ``$\mathrm{conn}$''.

The formulas for contributions to matrix elements
through the third-order of MBPT were presented in Ref.~\cite{BluGuoJoh87}.
The linearized coupled-cluster approach truncated at single and
double excitations (LCCSD) fully recovers the matrix elements through
the third order~\cite{BluJohLiu89}. Here we investigate the contributions
at the fourth order missed by the LCCSD method.

To derive the fourth-order correction to a matrix element,
we expand the matrix element and normalization factors
into powers of the residual Coulomb interaction
$Z_{wv} = \sum_{k=1} Z_{wv}^{(k)}$,
$N_{v} =  \sum_{k=0} N_v^{(k)}$. Further, we employ the
all-order result, Eq.~(\ref{Eqn_Zconn}), and expand the normalization
denominator into series. The result is
\begin{eqnarray}
M^{(4)}_{wv} &=&
\left\{
\langle \Psi_w^{(1)} |  Z  |\Psi_v^{(2)}  \rangle +
\langle \Psi_w^{(0)} | Z  |\Psi_v^{(3)}  \rangle + \right. \nonumber \\
& & \left.
\langle \Psi_w^{(2)} |  Z  |\Psi_v^{(1)}  \rangle +
\langle \Psi_w^{(3)} |  Z  |\Psi_v^{(0)}  \rangle \right\}_\mathrm{val,conn}
+  \nonumber \\
& &
+ Z^{(4)}_{wv, \, {\rm norm}}\label{Eq_Z4}     \, ,
\end{eqnarray}
where only connected valence contributions are to be kept. The
normalization correction is given by
\begin{eqnarray}
Z^{(4)}_{wv, \, {\rm norm}} &=&
- \frac{1}{2}\, \left( N_v^{(2)} + N_w^{(2)} \right)_\mathrm{val,conn}\,
  \left(Z_{wv}^{(2)}\right)_\mathrm{val,conn}  + \nonumber \\
& & - \frac{1}{2} \, \left( N_v^{(3)} + N_w^{(3)} \right)_\mathrm{val,conn} \, z_{wv} \, ,
\label{Eqn_Z4norm}
\end{eqnarray}
where we used that $N_v^{(1)}=0$ and $Z_{wv}^{(1)} \equiv z_{wv}$, the
matrix element in the Hartree-Fock approximation.

As we proceed to the derivation of the fourth-order
diagrams we notice that
the second line of Eq.~(\ref{Eq_Z4})
is the hermitian conjugate of the first line with a swap of valence indexes $w$ and $v$.
This observation allows us to consider only half of the diagrams
since in numerical evaluation the conjugated terms do not require additional
programming efforts.

\section{Discussion of fourth-order diagrams}

We fully derived the fourth-order correction to matrix elements using
Wick theorem. A set of simplification rules was implemented
with the symbolic algebra system {\em Mathematica}~\cite{Wol99}.
Excluding the normalization
correction and folded diagrams, the resulting number of diagrams in the fourth
order is 262. We counted both direct and all possible exchange forms of
a given diagram as a single contribution. We excluded hermitian
conjugated terms from the counting procedure.
The linearized coupled cluster approach, truncated at single and double
excitations (LCCSD) recovers approximately half of the fourth-order contributions.
The remaining diagrams are due to triple excitations (128 terms) and
nonlinear contribution of double excitations (14 terms). Explicit expressions
for these  complementary contributions are given in the Appendix.

\begin{figure}[h]
\begin{center}
\includegraphics*[scale=0.65]{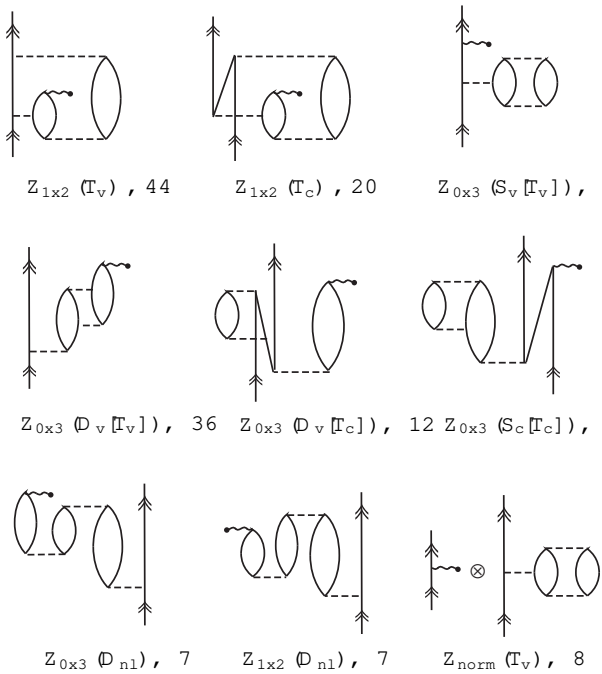}
\caption{ \small \label{Fig_Z4} Sample fourth-order diagrams involving triple
excitations and non-linear coupled-cluster contributions. The one-particle
matrix element is denoted by a wavy horizontal line.
See the explanation in the
text for diagram classification. The number of contributions
for each class of diagrams is also shown;
direct, all possible exchange, and the conjugated graphs of a given
diagram were counted as a single contribution.
}
\end{center}
\end{figure}

We break all fourth-order contributions complementary to the LCCSD subset of diagrams
into nine classes:
\begin{eqnarray}
\lefteqn{ \left( M^{(4)}_{wv} \right)_\mathrm{non-LCCSD} =
Z_{1 \times 2}(T_v) + Z_{1 \times 2}(T_c) + } \nonumber   \\
& &Z_{0 \times 3}(S_v[T_v]) + Z_{0 \times 3}(D_v[T_v]) +\\
& &Z_{0 \times 3}(S_c[T_c]) + Z_{0 \times 3}(D_v[T_c]) + \nonumber \\
& & Z_{1 \times 2}(D_{nl}) +  Z_{0 \times 3}(D_{nl})+
Z_\mathrm{norm}(T_v) \, . \nonumber
\end{eqnarray}
The representative
diagrams for each class of contributions are shown in Fig.~\ref{Fig_Z4}.
Here the diagrams $Z_{1 \times 2}(\ldots)$ arise from evaluation of
expression $\langle \Psi_w^{(1)} |  Z  |\Psi_v^{(2)}  \rangle $ and its hermitian
conjugate with a swap of valence labels $w$ and $v$.
Similarly $Z_{0 \times 3}(\ldots)$ terms are generated from
$\langle \Psi_w^{(0)} |  Z  |\Psi_v^{(3)}  \rangle +\mathrm{c.c.}$ Finally
$Z_\mathrm{norm}(\ldots)$ are due to normalization correction, Eq.~(\ref{Eqn_Z4norm}).
Further, we classify the diagrams by the presence of core triples ($T_c$) or
valence  triples ($T_v$). For $Z_{0 \times 3}(\ldots)$ terms triple excitations
occur as an intermediate contribution (see Fig.~\ref{Fig_Om3}) and
we distinguish the effect of triples on lower-rank excitations, e.g.
$D_v[T_c]$ is the effect of core triples on valence doubles. Finally, the diagrams
marked $D_{nl}$ are due to the effect of disconnected quadruple excitations.
These diagrams may be simplified to a direct product of double excitations,
as demonstrated in Fig.~\ref{Fig_Dnl}.

The introduced classes of diagrams are illustrated in Fig.~\ref{Fig_Z4}.
The numbers of contributions in each class are also given in that figure.
 Let us make some observations.
First of all, none of the diagrams contain
computationally intensive  Coulomb integrals involving
four particle states, e.g., $g_{mnrs}$.  We also notice
the absence of term $D_c[T_c]$, i.e., the effects of core triples on core double excitations.
The core triples also do not contribute to the normalization correction.
All these simplifications may lead to a design of an efficient numerical evaluation scheme.
%There is an equal number of contributions arising from $Z_{1 \times 2}(\ldots)$
%and $Z_{0 \times 3}(\ldots)$ products.

\begin{figure}[h]
\begin{center}
\includegraphics*[scale=0.65]{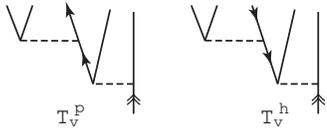}
\caption{ \small \label{Fig_Thp}
Separation of triple excitations based on a nature of an orbital line
connecting upper and lower interactions.
$T_v^p$ diagram involves particle line and $T_v^h$ a hole line. Similar separation
may be carried out for core triples. }
\end{center}
\end{figure}

The dominant number of diagrams is due
to valence triple excitations, the set $Z_{1 \times 2}(T_v)$ accounting for 44
and the set $Z_{0 \times 3}(D_v[T_v])$ for 36 contributions.
We further distinguish second-order triples $T$ by the nature of
the orbital line connecting upper and lower interactions $T=T^p+T^h$,
$T^p$ standing for a particle
line and $T^h$ for a hole line as illustrated in Fig.~\ref{Fig_Thp}.
Such a separation is motivated by considerations of computational complexity:
the $T^h$ diagram, involving summation over a small number of
core states, may be calculated much faster than a
similar $T^p$ contribution. We write
\begin{eqnarray*}
Z_{1 \times 2}(T_v)      &=& Z_{1 \times 2}(T_v^p) +  Z_{1 \times 2}(T_v^h)\, ,\\
Z_{0 \times 3}(D_v[T_v]) &=& Z_{0 \times 3}(D_v[T_v^p]) + Z_{0 \times 3}(D_v[T_v^h])\, .
\end{eqnarray*}
The formulas in the Appendix are grouped according to this scheme.

The effect of triples on {\em single} excitations in $\Omega_v^{(3)}$,
such as diagrams~\ref{Fig_Om3}(a) and (d), has been treated previously
in Refs.~\cite{BluJohSap91,SafJohDer99}.
Corresponding contributions to $M_{wv}^{(4)}$, $Z_{0 \times 3}(S_v[T_v])$ and
$Z_{0 \times 3}(S_c[T_c])$
are shown in Fig.~\ref{Fig_Z4}.
It was found that this effect contributes as much as 5\%  to
hyperfine-structure constants in Cs, and brings the {\em ab initio} calculations
into 0.5\% agreement
with experiment. At the same time the experiment-theory agreement
becomes worse for electric-dipole matrix elements when the $S[T]$
effect is included.
To fully understand the role of triple excitations it is important
to investigate all the enumerated effects on triple excitations, i.e.
effect of triples on valence doubles, direct contribution
of triple excitations to matrix elements entering  $Z_{1 \times 2}$,
and also  the normalization correction due to valence triple excitations.

The {\em linearized} coupled-cluster method (LCCSD)
\cite{BluJohLiu89,BluJohSap91,SafDerJoh98,SafJohDer99}
additionally disregards nonlinear terms in the coupled-cluster expansion. Therefore contributions
to  $\Omega_v^{(3)}$ similar to one shown in Fig.~\ref{Fig_Om3}(f) are omitted.
These nonlinear terms lead to additional corrections $Z_{0 \times 3}(D_{nl})$.
A similar effect omitted in the LCCSD approach is a direct
contribution of disconnected double excitations to matrix elements
represented by the diagrams of  $Z_{1 \times 2}(D_{nl})$ class.
It is worth noting that consideration of the nonlinear contributions is  key
for accounting for the full set of random-phase-approximation diagrams with
the CCSD method.

We further notice that in the framework
of traditional Rayleigh-Schr\"{o}dinger perturbation theory there are also contributions
from so-called folded diagrams, as discussed in Section~\ref{Sec_LDE}.
These folded diagrams originate from the second-order valence
energy correction (the first-order correction is zero
in the frozen-core Hartree-Fock basis).
Since both the CCSD  method and its linearized version fully
recover the second order energies~\cite{BluJohLiu89}, in our approach we have omitted
contributions of the folded diagrams.

Finally, we would like to comment on a possible all-order extension
of the derived fourth-order contributions.  Ideally, the entire
fourth order set of diagrams would be recovered by fully treating
the triple and nonlinear double  excitations  within the traditional coupled-cluster approach.
However, at the present state of computer technology such a full treatment hardly seems
feasible in relativistic calculations.  At the same time the coupled-cluster
expansion truncated at the single and double excitations (CCSD method) presents an
attractive starting point. The triple excitations may be treated semi-perturbatively,
i.e., the triple excitations are replaced by a combination of ``bare'' Coulomb interaction and
an all-order CCSD double excitation~\cite{BluJohSap91}.

The following modifications of the CCSD method
should be made to partially sum the derived diagrams to all orders of
perturbation theory:
(i) Four classes of the derived diagrams
($Z_{0 \times 3}(S_v[T_v]), Z_{0 \times 3}(D_v[T_v]),
Z_{0 \times 3}(S_c[T_c]), Z_{0 \times 3}(D_v[T_c])$) may be accounted
for by amending the traditional CCSD equations
with a semi-perturbative contribution of triple excitations.
Two of the desired modifications, $S_v[T_v]$ and $S_c[T_c]$,
were considered previously in Ref.~\cite{BluJohSap91,SafJohDer99}.
(ii) In the diagrams $Z_{1 \times 2}(T_v)$, $Z_{1 \times 2}(T_c)$, and $Z_\mathrm{norm}(T_v)$
the bottom and the upper (closing)  Coulomb interactions should be replaced by
all-order double excitation amplitudes.  This generalization follows from considering
the relevant contributions in the coupled-cluster method.
(iii) In the $Z_{1 \times 2}(D_{nl})$ diagrams all the Coulomb interactions should be
replaced by all-order double excitation amplitudes.
(iv) The linearized coupled-cluster
expansions should include terms nonlinear in double excitations to recover the
diagrams $Z_{0 \times 3}(D_{nl})$ in all-order fashion.

\section{Conclusion}
An improvement of the accuracy of {\em ab initio} Coulomb-correlated calculations
is necessitated by the latest experimental and theoretical  progress in
studies of parity violation in alkali-metal atoms.
Such improvement may  possibly be achieved by augmenting powerful all-order
techniques by contributions missed  in a given
order of many-body perturbation theory.
We derived and analyzed the entire set of fourth-order many-body diagrams for a one-particle
operator.

We highlighted the  fourth-order contributions
omitted in the popular coupled-cluster approach truncated
at single and double excitations (CCSD). To recover the full set of
fourth-order diagrams one should additionally consider the effect of triple
excitations. In addition, the linearized version of CCSD should
be augmented by nonlinear contributions of double excitations.
We presented explicit formulas for such complementary contributions in the Appendix.
The representative diagrams may be found in Fig.~\ref{Fig_Z4}. We also proposed a
possible extension of the derived fourth-order contributions to all orders of perturbation
theory.

The derived expressions may be useful for an analysis of the completeness of all-order methods
in the fourth order of perturbation theory and for designs of
next-generation approximations in atomic many-body calculations.

\begin{acknowledgments}
We would like to thank Sergey Porsev for comments on the manuscript.
This work was supported in part by the National Science Foundation.
\end{acknowledgments}

\appendix
\section{Fourth-order corrections to matrix elements}
Here we tabulate fourth-order corrections to matrix elements of one-particle operator
involving triple excitations and nonlinear contribution from double excitations.
The classification of the diagrams and notation were introduced in the main text
of the paper. Briefly, the matrix elements $g_{ijlk}$ of the Coulomb interaction
are defined by Eq.~(\ref{Eqn_CoulMel}). The quantities $\tilde{g}_{ijlk}$
are antisymmetric combinations $\tilde{g}_{ijlk}={g}_{ijlk}- {g}_{ijkl}$.
Matrix elements of a non-scalar one-particle operator $Z$ are denoted as
$z_{ij}$. Core
orbitals are enumerated by letters $a,b,c,d$,  complementary
excited states are labelled by $m,n,r,s$, and
valence orbitals are denoted by $v$ and $w$. The notation
$\varepsilon_{xy\ldots z}$  stands for
$\varepsilon_{x} + \varepsilon_{y} + \cdots \varepsilon_{z}$.
The terms denoted $c.c.$ are to be calculated by taking the hermitian conjugate
of all preceding contributions and swapping labels $v$ and $w$.

For convenience of drawing the graphs, the sequence of interactions
in numerators is sorted so that the interaction to the right of another
interaction appears lower in the corresponding Brueckner-Goldstone diagram.

\begin{eqnarray*}
\lefteqn{ Z_{1 \times 2}(T_c)=} \\
&& \sum_{abcmnr} \frac{{\tilde{g}}_{abnr}{z_{cv}}{g_{nrcm}}{g_{mwab}}}
    {({{\varepsilon }_{mw}}-{{\varepsilon }_{ab}})\,
      ({{\varepsilon }_{nr}}-{{\varepsilon }_{ab}})\,
      ({{\varepsilon }_{nrw}}-{{\varepsilon }_{abc}})}\,+ \\
&&\sum_{abcmnr} \frac{{\tilde{g}}_{abnr}{z_{cv}}{\tilde{g}}_{rwcm}{g_{mnab}}}
    {({{\varepsilon }_{mn}}-{{\varepsilon }_{ab}})\,
      ({{\varepsilon }_{nr}}-{{\varepsilon }_{ab}})\,
      ({{\varepsilon }_{nrw}}-{{\varepsilon }_{abc}})}\,+ \\
&&-\sum_{abcmnr} \frac{{\tilde{g}}_{abnv}{z_{cr}}{\tilde{g}}_{rwcm}{g_{mnab}}}
     {({{\varepsilon }_{mn}}-{{\varepsilon }_{ab}})\,
       ({{\varepsilon }_{nv}}-{{\varepsilon }_{ab}})\,
       ({{\varepsilon }_{nrw}}-{{\varepsilon }_{abc}})}\,+ \\
&&\sum_{abcmnr} \frac{{\tilde{g}}_{abrv}{z_{cn}}{\tilde{g}}_{nrcm}{g_{mwab}}}
    {({{\varepsilon }_{mw}}-{{\varepsilon }_{ab}})\,
      ({{\varepsilon }_{rv}}-{{\varepsilon }_{ab}})\,
      ({{\varepsilon }_{nrw}}-{{\varepsilon }_{abc}})}\,+
\\ &&
\sum_{abcmnr} \frac{{\tilde{g}}_{abrv}{z_{cn}}{\tilde{g}}_{rwcm}{g_{mnab}}}
    {({{\varepsilon }_{mn}}-{{\varepsilon }_{ab}})\,
      ({{\varepsilon }_{rv}}-{{\varepsilon }_{ab}})\,
      ({{\varepsilon }_{nrw}}-{{\varepsilon }_{abc}})}\,+
\\ &&
\sum_{abcmnr} \frac{{\tilde{g}}_{bcnr}{z_{av}}{g_{nrcm}}{\tilde{g}}_{mwab}}
    {({{\varepsilon }_{mw}}-{{\varepsilon }_{ab}})\,
      ({{\varepsilon }_{nr}}-{{\varepsilon }_{bc}})\,
      ({{\varepsilon }_{nrw}}-{{\varepsilon }_{abc}})}\,+
\\ &&
\sum_{abcmnr} \frac{{\tilde{g}}_{bcnr}{z_{av}}{\tilde{g}}_{rwcm}
      {\tilde{g}}_{mnab}}{({{\varepsilon }_{mn}}-{{\varepsilon }_{ab}})\,
      ({{\varepsilon }_{nr}}-{{\varepsilon }_{bc}})\,
      ({{\varepsilon }_{nrw}}-{{\varepsilon }_{abc}})}\,+
\\ &&
-\sum_{abcmnr} \frac{{\tilde{g}}_{bcnv}{z_{ar}}{\tilde{g}}_{rwcm}
       {\tilde{g}}_{mnab}}{({{\varepsilon }_{mn}}-{{\varepsilon }_{ab}})\,
       ({{\varepsilon }_{nv}}-{{\varepsilon }_{bc}})\,
       ({{\varepsilon }_{nrw}}-{{\varepsilon }_{abc}})}\,+
\\ &&
\sum_{abcmnr} \frac{{\tilde{g}}_{bcrv}{z_{an}}{\tilde{g}}_{nrcm}
      {\tilde{g}}_{mwab}}{({{\varepsilon }_{mw}}-{{\varepsilon }_{ab}})\,
      ({{\varepsilon }_{rv}}-{{\varepsilon }_{bc}})\,
      ({{\varepsilon }_{nrw}}-{{\varepsilon }_{abc}})}\,+
\\ &&
\sum_{abcmnr} \frac{{\tilde{g}}_{bcrv}{z_{an}}{\tilde{g}}_{rwcm}
      {\tilde{g}}_{mnab}}{({{\varepsilon }_{mn}}-{{\varepsilon }_{ab}})\,
      ({{\varepsilon }_{rv}}-{{\varepsilon }_{bc}})\,
      ({{\varepsilon }_{nrw}}-{{\varepsilon }_{abc}})}\,+
\\ &&
\sum_{abcdmn} \frac{{\tilde{g}}_{bdmn}{z_{cv}}{\tilde{g}}_{ancd}
      {\tilde{g}}_{mwab}}{({{\varepsilon }_{mn}}-{{\varepsilon }_{bd}})\,
      ({{\varepsilon }_{mw}}-{{\varepsilon }_{ab}})\,
      ({{\varepsilon }_{mnw}}-{{\varepsilon }_{bcd}})}\,+
\\ &&
-\sum_{abcdmn} \frac{{\tilde{g}}_{bdmn}{z_{cv}}{\tilde{g}}_{awcd}{g_{mnab}}}
     {({{\varepsilon }_{mn}}-{{\varepsilon }_{ab}})\,
       ({{\varepsilon }_{mn}}-{{\varepsilon }_{bd}})\,
       ({{\varepsilon }_{mnw}}-{{\varepsilon }_{bcd}})}\,+
\\ &&
-\sum_{abcdmn} \frac{{\tilde{g}}_{bdmv}{z_{cn}}{\tilde{g}}_{ancd}
       {\tilde{g}}_{mwab}}{({{\varepsilon }_{mv}}-{{\varepsilon }_{bd}})\,
       ({{\varepsilon }_{mw}}-{{\varepsilon }_{ab}})\,
       ({{\varepsilon }_{mnw}}-{{\varepsilon }_{bcd}})}\,+
\\ &&
\sum_{abcdmn} \frac{{\tilde{g}}_{bdnv}{z_{cm}}{\tilde{g}}_{ancd}
      {\tilde{g}}_{mwab}}{({{\varepsilon }_{mw}}-{{\varepsilon }_{ab}})\,
      ({{\varepsilon }_{nv}}-{{\varepsilon }_{bd}})\,
      ({{\varepsilon }_{mnw}}-{{\varepsilon }_{bcd}})}\,+
\\ &&
-\sum_{abcdmn} \frac{{\tilde{g}}_{bdnv}{z_{cm}}{\tilde{g}}_{awcd}
       {\tilde{g}}_{mnab}}{({{\varepsilon }_{mn}}-{{\varepsilon }_{ab}})\,
       ({{\varepsilon }_{nv}}-{{\varepsilon }_{bd}})\,
       ({{\varepsilon }_{mnw}}-{{\varepsilon }_{bcd}})}\,+
\\ &&
-\sum_{abcdmn} \frac{{\tilde{g}}_{cdmn}{z_{bv}}{g_{ancd}}{\tilde{g}}_{mwab}}
     {({{\varepsilon }_{mn}}-{{\varepsilon }_{cd}})\,
       ({{\varepsilon }_{mw}}-{{\varepsilon }_{ab}})\,
       ({{\varepsilon }_{mnw}}-{{\varepsilon }_{bcd}})}\,+
\\ &&
\sum_{abcdmn} \frac{{\tilde{g}}_{cdmn}{z_{bv}}{g_{awcd}}{g_{mnab}}}
    {({{\varepsilon }_{mn}}-{{\varepsilon }_{ab}})\,
      ({{\varepsilon }_{mn}}-{{\varepsilon }_{cd}})\,
      ({{\varepsilon }_{mnw}}-{{\varepsilon }_{bcd}})}\,+
\\ &&
\sum_{abcdmn} \frac{{\tilde{g}}_{cdmv}{z_{bn}}{g_{ancd}}{\tilde{g}}_{mwab}}
    {({{\varepsilon }_{mv}}-{{\varepsilon }_{cd}})\,
      ({{\varepsilon }_{mw}}-{{\varepsilon }_{ab}})\,
      ({{\varepsilon }_{mnw}}-{{\varepsilon }_{bcd}})}\,+
\\ &&
-\sum_{abcdmn} \frac{{\tilde{g}}_{cdnv}{z_{bm}}{g_{ancd}}{\tilde{g}}_{mwab}}
     {({{\varepsilon }_{mw}}-{{\varepsilon }_{ab}})\,
       ({{\varepsilon }_{nv}}-{{\varepsilon }_{cd}})\,
       ({{\varepsilon }_{mnw}}-{{\varepsilon }_{bcd}})}\,+
\\ &&
\sum_{abcdmn} \frac{{\tilde{g}}_{cdnv}{z_{bm}}{g_{awcd}}{\tilde{g}}_{mnab}}
    {({{\varepsilon }_{mn}}-{{\varepsilon }_{ab}})\,
      ({{\varepsilon }_{nv}}-{{\varepsilon }_{cd}})\,
      ({{\varepsilon }_{mnw}}-{{\varepsilon }_{bcd}})}\,
      + c.c.
\end{eqnarray*}

\begin{eqnarray*}
\lefteqn{Z_{1 \times 2}(T^h_v) =}\\
&&-\sum_{abcmnr} \frac{{\tilde{g}}_{bcmn}{z_{wr}}{g_{arbc}}{g_{mnav}}}
     {({{\varepsilon }_{mn}}-{{\varepsilon }_{av}})\,
       ({{\varepsilon }_{mn}}-{{\varepsilon }_{bc}})\,
       ({{\varepsilon }_{mnr}}-{{\varepsilon }_{bcv}})}\,+
\\ &&
-\sum_{abcmnr} \frac{{\tilde{g}}_{bcmn}{z_{wr}}{\tilde{g}}_{arcv}{g_{mnab}}}
     {({{\varepsilon }_{mn}}-{{\varepsilon }_{ab}})\,
       ({{\varepsilon }_{mn}}-{{\varepsilon }_{bc}})\,
       ({{\varepsilon }_{mnr}}-{{\varepsilon }_{bcv}})}\,+
\\ &&
\sum_{abcmnr} \frac{{\tilde{g}}_{bcmr}{z_{rn}}{g_{anbc}}{\tilde{g}}_{mwav}}
    {({{\varepsilon }_{mr}}-{{\varepsilon }_{bc}})\,
      ({{\varepsilon }_{mw}}-{{\varepsilon }_{av}})\,
      ({{\varepsilon }_{mnw}}-{{\varepsilon }_{bcv}})}\,+
\\ &&
\sum_{abcmnr} \frac{{\tilde{g}}_{bcmr}{z_{rn}}{\tilde{g}}_{ancv}
      {\tilde{g}}_{mwab}}{({{\varepsilon }_{mr}}-{{\varepsilon }_{bc}})\,
      ({{\varepsilon }_{mw}}-{{\varepsilon }_{ab}})\,
      ({{\varepsilon }_{mnw}}-{{\varepsilon }_{bcv}})}\,+
\\ &&
-\sum_{abcmnr} \frac{{\tilde{g}}_{bcnr}{z_{rm}}{g_{anbc}}{\tilde{g}}_{mwav}}
     {({{\varepsilon }_{mw}}-{{\varepsilon }_{av}})\,
       ({{\varepsilon }_{nr}}-{{\varepsilon }_{bc}})\,
       ({{\varepsilon }_{mnw}}-{{\varepsilon }_{bcv}})}\,+
\\ &&
\sum_{abcmnr} \frac{{\tilde{g}}_{bcnr}{z_{rm}}{g_{awbc}}{\tilde{g}}_{mnav}}
    {({{\varepsilon }_{mn}}-{{\varepsilon }_{av}})\,
      ({{\varepsilon }_{nr}}-{{\varepsilon }_{bc}})\,
      ({{\varepsilon }_{mnw}}-{{\varepsilon }_{bcv}})}\,+
\\ &&
-\sum_{abcmnr} \frac{{\tilde{g}}_{bcnr}{z_{rm}}{\tilde{g}}_{ancv}
       {\tilde{g}}_{mwab}}{({{\varepsilon }_{mw}}-{{\varepsilon }_{ab}})\,
       ({{\varepsilon }_{nr}}-{{\varepsilon }_{bc}})\,
       ({{\varepsilon }_{mnw}}-{{\varepsilon }_{bcv}})}\,+
\\ &&
\sum_{abcmnr} \frac{{\tilde{g}}_{bcnr}{z_{rm}}{\tilde{g}}_{awcv}
      {\tilde{g}}_{mnab}}{({{\varepsilon }_{mn}}-{{\varepsilon }_{ab}})\,
      ({{\varepsilon }_{nr}}-{{\varepsilon }_{bc}})\,
      ({{\varepsilon }_{mnw}}-{{\varepsilon }_{bcv}})}\,+
\\ &&
-\sum_{abcmnr} \frac{{\tilde{g}}_{bcnr}{z_{wm}}{g_{arbc}}{\tilde{g}}_{mnav}}
     {({{\varepsilon }_{mn}}-{{\varepsilon }_{av}})\,
       ({{\varepsilon }_{nr}}-{{\varepsilon }_{bc}})\,
       ({{\varepsilon }_{mnr}}-{{\varepsilon }_{bcv}})}\,+
\\ &&
-\sum_{abcmnr} \frac{{\tilde{g}}_{bcnr}{z_{wm}}{\tilde{g}}_{arcv}
       {\tilde{g}}_{mnab}}{({{\varepsilon }_{mn}}-{{\varepsilon }_{ab}})\,
       ({{\varepsilon }_{nr}}-{{\varepsilon }_{bc}})\,
       ({{\varepsilon }_{mnr}}-{{\varepsilon }_{bcv}})}\,+
\\ &&
-\sum_{abcdmn} \frac{{\tilde{g}}_{bdmn}{z_{cd}}{\tilde{g}}_{ancv}
       {\tilde{g}}_{mwab}}{({{\varepsilon }_{mn}}-{{\varepsilon }_{bd}})\,
       ({{\varepsilon }_{mw}}-{{\varepsilon }_{ab}})\,
       ({{\varepsilon }_{mnw}}-{{\varepsilon }_{bcv}})}\,+
\\ &&
\sum_{abcdmn} \frac{{\tilde{g}}_{bdmn}{z_{cd}}{\tilde{g}}_{awcv}{g_{mnab}}}
    {({{\varepsilon }_{mn}}-{{\varepsilon }_{ab}})\,
      ({{\varepsilon }_{mn}}-{{\varepsilon }_{bd}})\,
      ({{\varepsilon }_{mnw}}-{{\varepsilon }_{bcv}})}\,+
\\ &&
\sum_{abcmnr} \frac{{\tilde{g}}_{bwmn}{z_{cr}}{\tilde{g}}_{arcv}{g_{mnab}}}
    {({{\varepsilon }_{mn}}-{{\varepsilon }_{ab}})\,
      ({{\varepsilon }_{mn}}-{{\varepsilon }_{bw}})\,
      ({{\varepsilon }_{mnr}}-{{\varepsilon }_{bcv}})}\,+
\\ &&
\sum_{abcmnr} \frac{{\tilde{g}}_{bwnr}{z_{cm}}{\tilde{g}}_{arcv}
      {\tilde{g}}_{mnab}}{({{\varepsilon }_{mn}}-{{\varepsilon }_{ab}})\,
      ({{\varepsilon }_{nr}}-{{\varepsilon }_{bw}})\,
      ({{\varepsilon }_{mnr}}-{{\varepsilon }_{bcv}})}\,+
\\ &&
\sum_{abcdmn} \frac{{\tilde{g}}_{cdmn}{z_{bd}}{\tilde{g}}_{anbc}
      {\tilde{g}}_{mwav}}{({{\varepsilon }_{mn}}-{{\varepsilon }_{cd}})\,
      ({{\varepsilon }_{mw}}-{{\varepsilon }_{av}})\,
      ({{\varepsilon }_{mnw}}-{{\varepsilon }_{bcv}})}\,+
\\ &&
\sum_{abcdmn} \frac{{\tilde{g}}_{cdmn}{z_{bd}}{\tilde{g}}_{ancv}
      {\tilde{g}}_{mwab}}{({{\varepsilon }_{mn}}-{{\varepsilon }_{cd}})\,
      ({{\varepsilon }_{mw}}-{{\varepsilon }_{ab}})\,
      ({{\varepsilon }_{mnw}}-{{\varepsilon }_{bcv}})}\,+
\\ &&
-\sum_{abcdmn} \frac{{\tilde{g}}_{cdmn}{z_{bd}}{\tilde{g}}_{awbc}{g_{mnav}}}
     {({{\varepsilon }_{mn}}-{{\varepsilon }_{av}})\,
       ({{\varepsilon }_{mn}}-{{\varepsilon }_{cd}})\,
       ({{\varepsilon }_{mnw}}-{{\varepsilon }_{bcv}})}\,+
\\ &&
-\sum_{abcdmn} \frac{{\tilde{g}}_{cdmn}{z_{bd}}{\tilde{g}}_{awcv}{g_{mnab}}}
     {({{\varepsilon }_{mn}}-{{\varepsilon }_{ab}})\,
       ({{\varepsilon }_{mn}}-{{\varepsilon }_{cd}})\,
       ({{\varepsilon }_{mnw}}-{{\varepsilon }_{bcv}})}\,+
\\ &&
-\sum_{abcmnr} \frac{{\tilde{g}}_{cwmn}{z_{br}}{\tilde{g}}_{arbc}{g_{mnav}}}
     {({{\varepsilon }_{mn}}-{{\varepsilon }_{av}})\,
       ({{\varepsilon }_{mn}}-{{\varepsilon }_{cw}})\,
       ({{\varepsilon }_{mnr}}-{{\varepsilon }_{bcv}})}\,+
\\ &&
-\sum_{abcmnr} \frac{{\tilde{g}}_{cwmn}{z_{br}}{\tilde{g}}_{arcv}{g_{mnab}}}
     {({{\varepsilon }_{mn}}-{{\varepsilon }_{ab}})\,
       ({{\varepsilon }_{mn}}-{{\varepsilon }_{cw}})\,
       ({{\varepsilon }_{mnr}}-{{\varepsilon }_{bcv}})}\,+
\\ &&
-\sum_{abcmnr} \frac{{\tilde{g}}_{cwnr}{z_{bm}}{\tilde{g}}_{arbc}
       {\tilde{g}}_{mnav}}{({{\varepsilon }_{mn}}-{{\varepsilon }_{av}})\,
       ({{\varepsilon }_{nr}}-{{\varepsilon }_{cw}})\,
       ({{\varepsilon }_{mnr}}-{{\varepsilon }_{bcv}})}\,+
\\ &&
-\sum_{abcmnr} \frac{{\tilde{g}}_{cwnr}{z_{bm}}{\tilde{g}}_{arcv}
       {\tilde{g}}_{mnab}}{({{\varepsilon }_{mn}}-{{\varepsilon }_{ab}})\,
       ({{\varepsilon }_{nr}}-{{\varepsilon }_{cw}})\,
       ({{\varepsilon }_{mnr}}-{{\varepsilon }_{bcv}})}\,
             + c.c.
\end{eqnarray*}

\begin{eqnarray*}
\lefteqn{Z_{1 \times 2}(T^p_v) =}\\
&&\sum_{abmnrs} \frac{{\tilde{g}}_{abns}{z_{sr}}{\tilde{g}}_{rwbm}
      {\tilde{g}}_{mnav}}{({{\varepsilon }_{mn}}-{{\varepsilon }_{av}})\,
      ({{\varepsilon }_{ns}}-{{\varepsilon }_{ab}})\,
      ({{\varepsilon }_{nrw}}-{{\varepsilon }_{abv}})}\,+
\\ &&
\sum_{abmnrs} \frac{{\tilde{g}}_{abns}{z_{sr}}{\tilde{g}}_{rwmv}{g_{mnab}}}
    {({{\varepsilon }_{mn}}-{{\varepsilon }_{ab}})\,
      ({{\varepsilon }_{ns}}-{{\varepsilon }_{ab}})\,
      ({{\varepsilon }_{nrw}}-{{\varepsilon }_{abv}})}\,+
\\ &&
-\sum_{abmnrs} \frac{{\tilde{g}}_{abns}{z_{wr}}{\tilde{g}}_{rsbm}
       {\tilde{g}}_{mnav}}{({{\varepsilon }_{mn}}-{{\varepsilon }_{av}})\,
       ({{\varepsilon }_{ns}}-{{\varepsilon }_{ab}})\,
       ({{\varepsilon }_{nrs}}-{{\varepsilon }_{abv}})}\,+
\\ &&
-\sum_{abmnrs} \frac{{\tilde{g}}_{abns}{z_{wr}}{\tilde{g}}_{rsmv}{g_{mnab}}}
     {({{\varepsilon }_{mn}}-{{\varepsilon }_{ab}})\,
       ({{\varepsilon }_{ns}}-{{\varepsilon }_{ab}})\,
       ({{\varepsilon }_{nrs}}-{{\varepsilon }_{abv}})}\,+
\\ &&
-\sum_{abmnrs} \frac{{\tilde{g}}_{abrs}{z_{sn}}{\tilde{g}}_{nrbm}
       {\tilde{g}}_{mwav}}{({{\varepsilon }_{mw}}-{{\varepsilon }_{av}})\,
       ({{\varepsilon }_{rs}}-{{\varepsilon }_{ab}})\,
       ({{\varepsilon }_{nrw}}-{{\varepsilon }_{abv}})}\,+
\\ &&
-\sum_{abmnrs} \frac{{\tilde{g}}_{abrs}{z_{sn}}{\tilde{g}}_{nrmv}{g_{mwab}}}
     {({{\varepsilon }_{mw}}-{{\varepsilon }_{ab}})\,
       ({{\varepsilon }_{rs}}-{{\varepsilon }_{ab}})\,
       ({{\varepsilon }_{nrw}}-{{\varepsilon }_{abv}})}\,+
\\ &&
-\sum_{abmnrs} \frac{{\tilde{g}}_{abrs}{z_{sn}}{\tilde{g}}_{rwbm}
       {\tilde{g}}_{mnav}}{({{\varepsilon }_{mn}}-{{\varepsilon }_{av}})\,
       ({{\varepsilon }_{rs}}-{{\varepsilon }_{ab}})\,
       ({{\varepsilon }_{nrw}}-{{\varepsilon }_{abv}})}\,+
\\ &&
-\sum_{abmnrs} \frac{{\tilde{g}}_{abrs}{z_{sn}}{\tilde{g}}_{rwmv}{g_{mnab}}}
     {({{\varepsilon }_{mn}}-{{\varepsilon }_{ab}})\,
       ({{\varepsilon }_{rs}}-{{\varepsilon }_{ab}})\,
       ({{\varepsilon }_{nrw}}-{{\varepsilon }_{abv}})}\,+
\\ &&
\sum_{abmnrs} \frac{{\tilde{g}}_{abrs}{z_{wn}}{g_{rsbm}}{\tilde{g}}_{mnav}}
    {({{\varepsilon }_{mn}}-{{\varepsilon }_{av}})\,
      ({{\varepsilon }_{rs}}-{{\varepsilon }_{ab}})\,
      ({{\varepsilon }_{nrs}}-{{\varepsilon }_{abv}})}\,+
\\ &&
\sum_{abmnrs} \frac{{\tilde{g}}_{abrs}{z_{wn}}{g_{rsmv}}{g_{mnab}}}
    {({{\varepsilon }_{mn}}-{{\varepsilon }_{ab}})\,
      ({{\varepsilon }_{rs}}-{{\varepsilon }_{ab}})\,
      ({{\varepsilon }_{nrs}}-{{\varepsilon }_{abv}})}\,+
\\ &&
-\sum_{abcmnr} \frac{{\tilde{g}}_{acnr}{z_{bc}}{g_{nrbm}}{\tilde{g}}_{mwav}}
     {({{\varepsilon }_{mw}}-{{\varepsilon }_{av}})\,
       ({{\varepsilon }_{nr}}-{{\varepsilon }_{ac}})\,
       ({{\varepsilon }_{nrw}}-{{\varepsilon }_{abv}})}\,+
\\ &&
-\sum_{abcmnr} \frac{{\tilde{g}}_{acnr}{z_{bc}}{\tilde{g}}_{rwbm}
       {\tilde{g}}_{mnav}}{({{\varepsilon }_{mn}}-{{\varepsilon }_{av}})\,
       ({{\varepsilon }_{nr}}-{{\varepsilon }_{ac}})\,
       ({{\varepsilon }_{nrw}}-{{\varepsilon }_{abv}})}\,+
\\ &&
\sum_{abmnrs} \frac{{\tilde{g}}_{awns}{z_{br}}{\tilde{g}}_{rsbm}
      {\tilde{g}}_{mnav}}{({{\varepsilon }_{mn}}-{{\varepsilon }_{av}})\,
      ({{\varepsilon }_{ns}}-{{\varepsilon }_{aw}})\,
      ({{\varepsilon }_{nrs}}-{{\varepsilon }_{abv}})}\,+
\\ &&
-\sum_{abmnrs} \frac{{\tilde{g}}_{awrs}{z_{bn}}{g_{rsbm}}{\tilde{g}}_{mnav}}
     {({{\varepsilon }_{mn}}-{{\varepsilon }_{av}})\,
       ({{\varepsilon }_{rs}}-{{\varepsilon }_{aw}})\,
       ({{\varepsilon }_{nrs}}-{{\varepsilon }_{abv}})}\,+
\\ &&
\sum_{abcmnr} \frac{{\tilde{g}}_{bcnr}{z_{ac}}{g_{nrbm}}{\tilde{g}}_{mwav}}
    {({{\varepsilon }_{mw}}-{{\varepsilon }_{av}})\,
      ({{\varepsilon }_{nr}}-{{\varepsilon }_{bc}})\,
      ({{\varepsilon }_{nrw}}-{{\varepsilon }_{abv}})}\,+
\\ &&
\sum_{abcmnr} \frac{{\tilde{g}}_{bcnr}{z_{ac}}{g_{nrmv}}{\tilde{g}}_{mwab}}
    {({{\varepsilon }_{mw}}-{{\varepsilon }_{ab}})\,
      ({{\varepsilon }_{nr}}-{{\varepsilon }_{bc}})\,
      ({{\varepsilon }_{nrw}}-{{\varepsilon }_{abv}})}\,+
\\ &&
\sum_{abcmnr} \frac{{\tilde{g}}_{bcnr}{z_{ac}}{\tilde{g}}_{rwbm}
      {\tilde{g}}_{mnav}}{({{\varepsilon }_{mn}}-{{\varepsilon }_{av}})\,
      ({{\varepsilon }_{nr}}-{{\varepsilon }_{bc}})\,
      ({{\varepsilon }_{nrw}}-{{\varepsilon }_{abv}})}\,+
\\ &&
\sum_{abcmnr} \frac{{\tilde{g}}_{bcnr}{z_{ac}}{\tilde{g}}_{rwmv}
      {\tilde{g}}_{mnab}}{({{\varepsilon }_{mn}}-{{\varepsilon }_{ab}})\,
      ({{\varepsilon }_{nr}}-{{\varepsilon }_{bc}})\,
      ({{\varepsilon }_{nrw}}-{{\varepsilon }_{abv}})}\,+
\\ &&
-\sum_{abmnrs} \frac{{\tilde{g}}_{bwns}{z_{ar}}{\tilde{g}}_{rsbm}
       {\tilde{g}}_{mnav}}{({{\varepsilon }_{mn}}-{{\varepsilon }_{av}})\,
       ({{\varepsilon }_{ns}}-{{\varepsilon }_{bw}})\,
       ({{\varepsilon }_{nrs}}-{{\varepsilon }_{abv}})}\,+
\\ &&
-\sum_{abmnrs} \frac{{\tilde{g}}_{bwns}{z_{ar}}{\tilde{g}}_{rsmv}
       {\tilde{g}}_{mnab}}{({{\varepsilon }_{mn}}-{{\varepsilon }_{ab}})\,
       ({{\varepsilon }_{ns}}-{{\varepsilon }_{bw}})\,
       ({{\varepsilon }_{nrs}}-{{\varepsilon }_{abv}})}\,+
\\ &&
\sum_{abmnrs} \frac{{\tilde{g}}_{bwrs}{z_{an}}{g_{rsbm}}{\tilde{g}}_{mnav}}
    {({{\varepsilon }_{mn}}-{{\varepsilon }_{av}})\,
      ({{\varepsilon }_{rs}}-{{\varepsilon }_{bw}})\,
      ({{\varepsilon }_{nrs}}-{{\varepsilon }_{abv}})}\,+
\\ &&
\sum_{abmnrs} \frac{{\tilde{g}}_{bwrs}{z_{an}}{g_{rsmv}}{\tilde{g}}_{mnab}}
    {({{\varepsilon }_{mn}}-{{\varepsilon }_{ab}})\,
      ({{\varepsilon }_{rs}}-{{\varepsilon }_{bw}})\,
      ({{\varepsilon }_{nrs}}-{{\varepsilon }_{abv}})}\,
            + c.c.
\end{eqnarray*}

\begin{eqnarray*}
\lefteqn{Z_{0 \times 3}(S_v[T_v]) =} \\
&&
\sum_{abmnrs} \frac{{z_{wn}}{\tilde{g}}_{abrs}{g_{rsbm}}{\tilde{g}}_{mnav}}
    {\left( {{\varepsilon }_n} - {{\varepsilon }_{v}} \right) \,
      ({{\varepsilon }_{mn}}-{{\varepsilon }_{av}})\,
      ({{\varepsilon }_{nrs}}-{{\varepsilon }_{abv}})}\,+
\\ &&
\sum_{abmnrs} \frac{{z_{wn}}{\tilde{g}}_{abrs}{g_{rsmv}}{g_{mnab}}}
    {\left( {{\varepsilon }_n} - {{\varepsilon }_{v}} \right) \,
      ({{\varepsilon }_{mn}}-{{\varepsilon }_{ab}})\,
      ({{\varepsilon }_{nrs}}-{{\varepsilon }_{abv}})}\,+
\\ &&
\sum_{abcmnr} \frac{{z_{wn}}{\tilde{g}}_{bcmr}{g_{arbc}}{\tilde{g}}_{mnav}}
    {\left( {{\varepsilon }_n} - {{\varepsilon }_{v}} \right) \,
      ({{\varepsilon }_{mn}}-{{\varepsilon }_{av}})\,
      ({{\varepsilon }_{mnr}}-{{\varepsilon }_{bcv}})}\,+
\\ &&
\sum_{abcmnr} \frac{{z_{wn}}{\tilde{g}}_{bcmr}{\tilde{g}}_{arcv}
      {\tilde{g}}_{mnab}}{\left( {{\varepsilon }_n} -
        {{\varepsilon }_{v}} \right) \,
      ({{\varepsilon }_{mn}}-{{\varepsilon }_{ab}})\,
      ({{\varepsilon }_{mnr}}-{{\varepsilon }_{bcv}})}\,+
\\ &&
-\sum_{abcmnr} \frac{{z_{wr}}{\tilde{g}}_{bcmn}{g_{arbc}}{g_{mnav}}}
     {\left( {{\varepsilon }_r} - {{\varepsilon }_{v}} \right) \,
       ({{\varepsilon }_{mn}}-{{\varepsilon }_{av}})\,
       ({{\varepsilon }_{mnr}}-{{\varepsilon }_{bcv}})}\,+
\\ &&
-\sum_{abcmnr} \frac{{z_{wr}}{\tilde{g}}_{bcmn}{\tilde{g}}_{arcv}{g_{mnab}}}
     {\left( {{\varepsilon }_r} - {{\varepsilon }_{v}} \right) \,
       ({{\varepsilon }_{mn}}-{{\varepsilon }_{ab}})\,
       ({{\varepsilon }_{mnr}}-{{\varepsilon }_{bcv}})}\,+
\\ &&
\sum_{abmnrs} \frac{{z_{ws}}{\tilde{g}}_{abnr}{\tilde{g}}_{rsbm}
      {\tilde{g}}_{mnav}}{\left( {{\varepsilon }_s} -
        {{\varepsilon }_{v}} \right) \,
      ({{\varepsilon }_{mn}}-{{\varepsilon }_{av}})\,
      ({{\varepsilon }_{nrs}}-{{\varepsilon }_{abv}})}\,+
\\ &&
\sum_{abmnrs} \frac{{z_{ws}}{\tilde{g}}_{abnr}{\tilde{g}}_{rsmv}{g_{mnab}}}
    {\left( {{\varepsilon }_s} - {{\varepsilon }_{v}} \right) \,
      ({{\varepsilon }_{mn}}-{{\varepsilon }_{ab}})\,
      ({{\varepsilon }_{nrs}}-{{\varepsilon }_{abv}})}\,
            + c.c.
\end{eqnarray*}

\begin{eqnarray*}
\lefteqn{Z_{0 \times 3}(S_c[T_c])=} \\
&&
-\sum_{abcmnr} \frac{{z_{bv}}{\tilde{g}}_{acnr}{g_{nrcm}}{\tilde{g}}_{mwab}}
     {({{\varepsilon }_w}-{{\varepsilon }_{b}})\,
       ({{\varepsilon }_{mw}}-{{\varepsilon }_{ab}})\,
       ({{\varepsilon }_{nrw}}-{{\varepsilon }_{abc}})}\,+
\\ &&
-\sum_{abcmnr} \frac{{z_{bv}}{\tilde{g}}_{acnr}{\tilde{g}}_{rwcm}
       {\tilde{g}}_{mnab}}{({{\varepsilon }_w}-{{\varepsilon }_{b}})\,
       ({{\varepsilon }_{mn}}-{{\varepsilon }_{ab}})\,
       ({{\varepsilon }_{nrw}}-{{\varepsilon }_{abc}})}\,+
\\ &&
-\sum_{abcdmn} \frac{{z_{bv}}{\tilde{g}}_{cdmn}{g_{ancd}}{\tilde{g}}_{mwab}}
     {({{\varepsilon }_w}-{{\varepsilon }_{b}})\,
       ({{\varepsilon }_{mw}}-{{\varepsilon }_{ab}})\,
       ({{\varepsilon }_{mnw}}-{{\varepsilon }_{bcd}})}\,+
\\ &&
\sum_{abcdmn} \frac{{z_{bv}}{\tilde{g}}_{cdmn}{g_{awcd}}{g_{mnab}}}
    {({{\varepsilon }_w}-{{\varepsilon }_{b}})\,
      ({{\varepsilon }_{mn}}-{{\varepsilon }_{ab}})\,
      ({{\varepsilon }_{mnw}}-{{\varepsilon }_{bcd}})}\,+
\\ &&
\sum_{abcmnr} \frac{{z_{cv}}{\tilde{g}}_{abnr}{g_{nrcm}}{g_{mwab}}}
    {({{\varepsilon }_w}-{{\varepsilon }_{c}})\,
      ({{\varepsilon }_{mw}}-{{\varepsilon }_{ab}})\,
      ({{\varepsilon }_{nrw}}-{{\varepsilon }_{abc}})}\,+
\\ &&
\sum_{abcmnr} \frac{{z_{cv}}{\tilde{g}}_{abnr}{\tilde{g}}_{rwcm}{g_{mnab}}}
    {({{\varepsilon }_w}-{{\varepsilon }_{c}})\,
      ({{\varepsilon }_{mn}}-{{\varepsilon }_{ab}})\,
      ({{\varepsilon }_{nrw}}-{{\varepsilon }_{abc}})}\,+
\\ &&
-\sum_{abcdmn} \frac{{z_{dv}}{\tilde{g}}_{bcmn}{\tilde{g}}_{ancd}
       {\tilde{g}}_{mwab}}{({{\varepsilon }_w}-{{\varepsilon }_{d}})\,
       ({{\varepsilon }_{mw}}-{{\varepsilon }_{ab}})\,
       ({{\varepsilon }_{mnw}}-{{\varepsilon }_{bcd}})}\,+
\\ &&
\sum_{abcdmn} \frac{{z_{dv}}{\tilde{g}}_{bcmn}{\tilde{g}}_{awcd}{g_{mnab}}}
    {({{\varepsilon }_w}-{{\varepsilon }_{d}})\,
      ({{\varepsilon }_{mn}}-{{\varepsilon }_{ab}})\,
      ({{\varepsilon }_{mnw}}-{{\varepsilon }_{bcd}})}\,
            + c.c.
\end{eqnarray*}

\begin{eqnarray*}
\lefteqn{Z_{0 \times 3}(D_v[T_c])=} \\
&&
-\sum_{abcdmn} \frac{{z_{bm}}{\tilde{g}}_{cdnv}{g_{ancd}}{\tilde{g}}_{mwab}}
     {({{\varepsilon }_{mw}}-{{\varepsilon }_{ab}})\,
       ({{\varepsilon }_{mw}}-{{\varepsilon }_{bv}})\,
       ({{\varepsilon }_{mnw}}-{{\varepsilon }_{bcd}})}\,+
\\ &&
-\sum_{abcmnr} \frac{{z_{bn}}{\tilde{g}}_{acrv}{\tilde{g}}_{rwcm}
       {\tilde{g}}_{mnab}}{({{\varepsilon }_{mn}}-{{\varepsilon }_{ab}})\,
       ({{\varepsilon }_{nw}}-{{\varepsilon }_{bv}})\,
       ({{\varepsilon }_{nrw}}-{{\varepsilon }_{abc}})}\,+
\\ &&
\sum_{abcdmn} \frac{{z_{bn}}{\tilde{g}}_{cdmv}{g_{ancd}}{\tilde{g}}_{mwab}}
    {({{\varepsilon }_{mw}}-{{\varepsilon }_{ab}})\,
      ({{\varepsilon }_{nw}}-{{\varepsilon }_{bv}})\,
      ({{\varepsilon }_{mnw}}-{{\varepsilon }_{bcd}})}\,+
\\ &&
-\sum_{abcdmn} \frac{{z_{bn}}{\tilde{g}}_{cdmv}{g_{awcd}}{\tilde{g}}_{mnab}}
     {({{\varepsilon }_{mn}}-{{\varepsilon }_{ab}})\,
       ({{\varepsilon }_{nw}}-{{\varepsilon }_{bv}})\,
       ({{\varepsilon }_{mnw}}-{{\varepsilon }_{bcd}})}\,+
\\ &&
\sum_{abcmnr} \frac{{z_{br}}{\tilde{g}}_{acnv}{\tilde{g}}_{nrcm}
      {\tilde{g}}_{mwab}}{({{\varepsilon }_{mw}}-{{\varepsilon }_{ab}})\,
      ({{\varepsilon }_{rw}}-{{\varepsilon }_{bv}})\,
      ({{\varepsilon }_{nrw}}-{{\varepsilon }_{abc}})}\,+
\\ &&
\sum_{abcmnr} \frac{{z_{br}}{\tilde{g}}_{acnv}{\tilde{g}}_{rwcm}
      {\tilde{g}}_{mnab}}{({{\varepsilon }_{mn}}-{{\varepsilon }_{ab}})\,
      ({{\varepsilon }_{rw}}-{{\varepsilon }_{bv}})\,
      ({{\varepsilon }_{nrw}}-{{\varepsilon }_{abc}})}\,+
\\ &&
\sum_{abcmnr} \frac{{z_{cn}}{\tilde{g}}_{abrv}{\tilde{g}}_{rwcm}{g_{mnab}}}
    {({{\varepsilon }_{mn}}-{{\varepsilon }_{ab}})\,
      ({{\varepsilon }_{nw}}-{{\varepsilon }_{cv}})\,
      ({{\varepsilon }_{nrw}}-{{\varepsilon }_{abc}})}\,+
\\ &&
-\sum_{abcmnr} \frac{{z_{cr}}{\tilde{g}}_{abnv}{\tilde{g}}_{nrcm}{g_{mwab}}}
     {({{\varepsilon }_{mw}}-{{\varepsilon }_{ab}})\,
       ({{\varepsilon }_{rw}}-{{\varepsilon }_{cv}})\,
       ({{\varepsilon }_{nrw}}-{{\varepsilon }_{abc}})}\,+
\\ &&
-\sum_{abcmnr} \frac{{z_{cr}}{\tilde{g}}_{abnv}{\tilde{g}}_{rwcm}{g_{mnab}}}
     {({{\varepsilon }_{mn}}-{{\varepsilon }_{ab}})\,
       ({{\varepsilon }_{rw}}-{{\varepsilon }_{cv}})\,
       ({{\varepsilon }_{nrw}}-{{\varepsilon }_{abc}})}\,+
\\ &&
-\sum_{abcdmn} \frac{{z_{dm}}{\tilde{g}}_{bcnv}{\tilde{g}}_{ancd}
       {\tilde{g}}_{mwab}}{({{\varepsilon }_{mw}}-{{\varepsilon }_{ab}})\,
       ({{\varepsilon }_{mw}}-{{\varepsilon }_{dv}})\,
       ({{\varepsilon }_{mnw}}-{{\varepsilon }_{bcd}})}\,+
\\ &&
\sum_{abcdmn} \frac{{z_{dn}}{\tilde{g}}_{bcmv}{\tilde{g}}_{ancd}
      {\tilde{g}}_{mwab}}{({{\varepsilon }_{mw}}-{{\varepsilon }_{ab}})\,
      ({{\varepsilon }_{nw}}-{{\varepsilon }_{dv}})\,
      ({{\varepsilon }_{mnw}}-{{\varepsilon }_{bcd}})}\,+
\\ &&
-\sum_{abcdmn} \frac{{z_{dn}}{\tilde{g}}_{bcmv}{\tilde{g}}_{awcd}
       {\tilde{g}}_{mnab}}{({{\varepsilon }_{mn}}-{{\varepsilon }_{ab}})\,
       ({{\varepsilon }_{nw}}-{{\varepsilon }_{dv}})\,
       ({{\varepsilon }_{mnw}}-{{\varepsilon }_{bcd}})}\,
             + c.c.
\end{eqnarray*}

\begin{eqnarray*}
\lefteqn{Z_{0 \times 3}(D_v[T_v^h])=} \\
&&
\sum_{abcmnr} \frac{{z_{bn}}{\tilde{g}}_{cwmr}{\tilde{g}}_{arcv}
      {\tilde{g}}_{mnab}}{({{\varepsilon }_{mn}}-{{\varepsilon }_{ab}})\,
      ({{\varepsilon }_{nw}}-{{\varepsilon }_{bv}})\,
      ({{\varepsilon }_{mnr}}-{{\varepsilon }_{bcv}})}\,+
\\ &&
-\sum_{abcmnr} \frac{{z_{br}}{\tilde{g}}_{crmn}{\tilde{g}}_{ancv}
       {\tilde{g}}_{mwab}}{({{\varepsilon }_{mw}}-{{\varepsilon }_{ab}})\,
       ({{\varepsilon }_{rw}}-{{\varepsilon }_{bv}})\,
       ({{\varepsilon }_{mnw}}-{{\varepsilon }_{bcv}})}\,+
\\ &&
\sum_{abcmnr} \frac{{z_{br}}{\tilde{g}}_{crmn}{\tilde{g}}_{awcv}{g_{mnab}}}
    {({{\varepsilon }_{mn}}-{{\varepsilon }_{ab}})\,
      ({{\varepsilon }_{rw}}-{{\varepsilon }_{bv}})\,
      ({{\varepsilon }_{mnw}}-{{\varepsilon }_{bcv}})}\,+
\\ &&
-\sum_{abcmnr} \frac{{z_{br}}{\tilde{g}}_{cwmn}{\tilde{g}}_{arcv}{g_{mnab}}}
     {({{\varepsilon }_{mn}}-{{\varepsilon }_{ab}})\,
       ({{\varepsilon }_{rw}}-{{\varepsilon }_{bv}})\,
       ({{\varepsilon }_{mnr}}-{{\varepsilon }_{bcv}})}\,+
\\ &&
-\sum_{abcmnr} \frac{{z_{cn}}{\tilde{g}}_{bwmr}{\tilde{g}}_{arbc}
       {\tilde{g}}_{mnav}}{({{\varepsilon }_{mn}}-{{\varepsilon }_{av}})\,
       ({{\varepsilon }_{nw}}-{{\varepsilon }_{cv}})\,
       ({{\varepsilon }_{mnr}}-{{\varepsilon }_{bcv}})}\,+
\\ &&
-\sum_{abcmnr} \frac{{z_{cn}}{\tilde{g}}_{bwmr}{\tilde{g}}_{arcv}
       {\tilde{g}}_{mnab}}{({{\varepsilon }_{mn}}-{{\varepsilon }_{ab}})\,
       ({{\varepsilon }_{nw}}-{{\varepsilon }_{cv}})\,
       ({{\varepsilon }_{mnr}}-{{\varepsilon }_{bcv}})}\,+
\\ &&
\sum_{abcmnr} \frac{{z_{cr}}{\tilde{g}}_{brmn}{\tilde{g}}_{anbc}
      {\tilde{g}}_{mwav}}{({{\varepsilon }_{mw}}-{{\varepsilon }_{av}})\,
      ({{\varepsilon }_{rw}}-{{\varepsilon }_{cv}})\,
      ({{\varepsilon }_{mnw}}-{{\varepsilon }_{bcv}})}\,+
\\ &&
\sum_{abcmnr} \frac{{z_{cr}}{\tilde{g}}_{brmn}{\tilde{g}}_{ancv}
      {\tilde{g}}_{mwab}}{({{\varepsilon }_{mw}}-{{\varepsilon }_{ab}})\,
      ({{\varepsilon }_{rw}}-{{\varepsilon }_{cv}})\,
      ({{\varepsilon }_{mnw}}-{{\varepsilon }_{bcv}})}\,+
\\ &&
-\sum_{abcmnr} \frac{{z_{cr}}{\tilde{g}}_{brmn}{\tilde{g}}_{awbc}{g_{mnav}}}
     {({{\varepsilon }_{mn}}-{{\varepsilon }_{av}})\,
       ({{\varepsilon }_{rw}}-{{\varepsilon }_{cv}})\,
       ({{\varepsilon }_{mnw}}-{{\varepsilon }_{bcv}})}\,+
\\ &&
-\sum_{abcmnr} \frac{{z_{cr}}{\tilde{g}}_{brmn}{\tilde{g}}_{awcv}{g_{mnab}}}
     {({{\varepsilon }_{mn}}-{{\varepsilon }_{ab}})\,
       ({{\varepsilon }_{rw}}-{{\varepsilon }_{cv}})\,
       ({{\varepsilon }_{mnw}}-{{\varepsilon }_{bcv}})}\,+
\\ &&
\sum_{abcmnr} \frac{{z_{cr}}{\tilde{g}}_{bwmn}{\tilde{g}}_{arbc}{g_{mnav}}}
    {({{\varepsilon }_{mn}}-{{\varepsilon }_{av}})\,
      ({{\varepsilon }_{rw}}-{{\varepsilon }_{cv}})\,
      ({{\varepsilon }_{mnr}}-{{\varepsilon }_{bcv}})}\,+
\\ &&
\sum_{abcmnr} \frac{{z_{cr}}{\tilde{g}}_{bwmn}{\tilde{g}}_{arcv}{g_{mnab}}}
    {({{\varepsilon }_{mn}}-{{\varepsilon }_{ab}})\,
      ({{\varepsilon }_{rw}}-{{\varepsilon }_{cv}})\,
      ({{\varepsilon }_{mnr}}-{{\varepsilon }_{bcv}})}\,+
\\ &&
-\sum_{abcdmn} \frac{{z_{dm}}{\tilde{g}}_{bcdn}{g_{anbc}}{\tilde{g}}_{mwav}}
     {({{\varepsilon }_{mw}}-{{\varepsilon }_{av}})\,
       ({{\varepsilon }_{mw}}-{{\varepsilon }_{dv}})\,
       ({{\varepsilon }_{mnw}}-{{\varepsilon }_{bcv}})}\,+
\\ &&
-\sum_{abcdmn} \frac{{z_{dm}}{\tilde{g}}_{bcdn}{\tilde{g}}_{ancv}
       {\tilde{g}}_{mwab}}{({{\varepsilon }_{mw}}-{{\varepsilon }_{ab}})\,
       ({{\varepsilon }_{mw}}-{{\varepsilon }_{dv}})\,
       ({{\varepsilon }_{mnw}}-{{\varepsilon }_{bcv}})}\,+
\\ &&
\sum_{abcdmn} \frac{{z_{dn}}{\tilde{g}}_{bcdm}{g_{anbc}}{\tilde{g}}_{mwav}}
    {({{\varepsilon }_{mw}}-{{\varepsilon }_{av}})\,
      ({{\varepsilon }_{nw}}-{{\varepsilon }_{dv}})\,
      ({{\varepsilon }_{mnw}}-{{\varepsilon }_{bcv}})}\,+
\\ &&
-\sum_{abcdmn} \frac{{z_{dn}}{\tilde{g}}_{bcdm}{g_{awbc}}{\tilde{g}}_{mnav}}
     {({{\varepsilon }_{mn}}-{{\varepsilon }_{av}})\,
       ({{\varepsilon }_{nw}}-{{\varepsilon }_{dv}})\,
       ({{\varepsilon }_{mnw}}-{{\varepsilon }_{bcv}})}\,+
\\ &&
\sum_{abcdmn} \frac{{z_{dn}}{\tilde{g}}_{bcdm}{\tilde{g}}_{ancv}
      {\tilde{g}}_{mwab}}{({{\varepsilon }_{mw}}-{{\varepsilon }_{ab}})\,
      ({{\varepsilon }_{nw}}-{{\varepsilon }_{dv}})\,
      ({{\varepsilon }_{mnw}}-{{\varepsilon }_{bcv}})}\,+
\\ &&
-\sum_{abcdmn} \frac{{z_{dn}}{\tilde{g}}_{bcdm}{\tilde{g}}_{awcv}
       {\tilde{g}}_{mnab}}{({{\varepsilon }_{mn}}-{{\varepsilon }_{ab}})\,
       ({{\varepsilon }_{nw}}-{{\varepsilon }_{dv}})\,
       ({{\varepsilon }_{mnw}}-{{\varepsilon }_{bcv}})}\,
             + c.c.
\end{eqnarray*}

\begin{eqnarray*}
\lefteqn{Z_{0 \times 3}(D_v[T_v^p])=} \\
&&
\sum_{abmnrs} \frac{{z_{an}}{\tilde{g}}_{bwrs}{g_{rsbm}}{\tilde{g}}_{mnav}}
    {({{\varepsilon }_{mn}}-{{\varepsilon }_{av}})\,
      ({{\varepsilon }_{nw}}-{{\varepsilon }_{av}})\,
      ({{\varepsilon }_{nrs}}-{{\varepsilon }_{abv}})}\,+
\\ &&
-\sum_{abmnrs} \frac{{z_{as}}{\tilde{g}}_{bsnr}{g_{nrbm}}{\tilde{g}}_{mwav}}
     {({{\varepsilon }_{mw}}-{{\varepsilon }_{av}})\,
       ({{\varepsilon }_{sw}}-{{\varepsilon }_{av}})\,
       ({{\varepsilon }_{nrw}}-{{\varepsilon }_{abv}})}\,+
\\ &&
-\sum_{abmnrs} \frac{{z_{as}}{\tilde{g}}_{bsnr}{\tilde{g}}_{rwbm}
       {\tilde{g}}_{mnav}}{({{\varepsilon }_{mn}}-{{\varepsilon }_{av}})\,
       ({{\varepsilon }_{sw}}-{{\varepsilon }_{av}})\,
       ({{\varepsilon }_{nrw}}-{{\varepsilon }_{abv}})}\,+
\\ &&
\sum_{abmnrs} \frac{{z_{as}}{\tilde{g}}_{bwnr}{\tilde{g}}_{rsbm}
      {\tilde{g}}_{mnav}}{({{\varepsilon }_{mn}}-{{\varepsilon }_{av}})\,
      ({{\varepsilon }_{sw}}-{{\varepsilon }_{av}})\,
      ({{\varepsilon }_{nrs}}-{{\varepsilon }_{abv}})}\,+
\\ &&
-\sum_{abmnrs} \frac{{z_{bn}}{\tilde{g}}_{awrs}{g_{rsbm}}{\tilde{g}}_{mnav}}
     {({{\varepsilon }_{mn}}-{{\varepsilon }_{av}})\,
       ({{\varepsilon }_{nw}}-{{\varepsilon }_{bv}})\,
       ({{\varepsilon }_{nrs}}-{{\varepsilon }_{abv}})}\,+
\\ &&
-\sum_{abmnrs} \frac{{z_{bn}}{\tilde{g}}_{awrs}{g_{rsmv}}{\tilde{g}}_{mnab}}
     {({{\varepsilon }_{mn}}-{{\varepsilon }_{ab}})\,
       ({{\varepsilon }_{nw}}-{{\varepsilon }_{bv}})\,
       ({{\varepsilon }_{nrs}}-{{\varepsilon }_{abv}})}\,+
\\ &&
\sum_{abmnrs} \frac{{z_{bs}}{\tilde{g}}_{asnr}{g_{nrbm}}{\tilde{g}}_{mwav}}
    {({{\varepsilon }_{mw}}-{{\varepsilon }_{av}})\,
      ({{\varepsilon }_{sw}}-{{\varepsilon }_{bv}})\,
      ({{\varepsilon }_{nrw}}-{{\varepsilon }_{abv}})}\,+
\\ &&
\sum_{abmnrs} \frac{{z_{bs}}{\tilde{g}}_{asnr}{g_{nrmv}}{\tilde{g}}_{mwab}}
    {({{\varepsilon }_{mw}}-{{\varepsilon }_{ab}})\,
      ({{\varepsilon }_{sw}}-{{\varepsilon }_{bv}})\,
      ({{\varepsilon }_{nrw}}-{{\varepsilon }_{abv}})}\,+
\\ &&
\sum_{abmnrs} \frac{{z_{bs}}{\tilde{g}}_{asnr}{\tilde{g}}_{rwbm}
      {\tilde{g}}_{mnav}}{({{\varepsilon }_{mn}}-{{\varepsilon }_{av}})\,
      ({{\varepsilon }_{sw}}-{{\varepsilon }_{bv}})\,
      ({{\varepsilon }_{nrw}}-{{\varepsilon }_{abv}})}\,+
\\ &&
\sum_{abmnrs} \frac{{z_{bs}}{\tilde{g}}_{asnr}{\tilde{g}}_{rwmv}
      {\tilde{g}}_{mnab}}{({{\varepsilon }_{mn}}-{{\varepsilon }_{ab}})\,
      ({{\varepsilon }_{sw}}-{{\varepsilon }_{bv}})\,
      ({{\varepsilon }_{nrw}}-{{\varepsilon }_{abv}})}\,+
\\ &&
-\sum_{abmnrs} \frac{{z_{bs}}{\tilde{g}}_{awnr}{\tilde{g}}_{rsbm}
       {\tilde{g}}_{mnav}}{({{\varepsilon }_{mn}}-{{\varepsilon }_{av}})\,
       ({{\varepsilon }_{sw}}-{{\varepsilon }_{bv}})\,
       ({{\varepsilon }_{nrs}}-{{\varepsilon }_{abv}})}\,+
\\ &&
-\sum_{abmnrs} \frac{{z_{bs}}{\tilde{g}}_{awnr}{\tilde{g}}_{rsmv}
       {\tilde{g}}_{mnab}}{({{\varepsilon }_{mn}}-{{\varepsilon }_{ab}})\,
       ({{\varepsilon }_{sw}}-{{\varepsilon }_{bv}})\,
       ({{\varepsilon }_{nrs}}-{{\varepsilon }_{abv}})}\,+
\\ &&
-\sum_{abcmnr} \frac{{z_{cn}}{\tilde{g}}_{abcr}{\tilde{g}}_{rwbm}
       {\tilde{g}}_{mnav}}{({{\varepsilon }_{mn}}-{{\varepsilon }_{av}})\,
       ({{\varepsilon }_{nw}}-{{\varepsilon }_{cv}})\,
       ({{\varepsilon }_{nrw}}-{{\varepsilon }_{abv}})}\,+
\\ &&
-\sum_{abcmnr} \frac{{z_{cn}}{\tilde{g}}_{abcr}{\tilde{g}}_{rwmv}{g_{mnab}}}
     {({{\varepsilon }_{mn}}-{{\varepsilon }_{ab}})\,
       ({{\varepsilon }_{nw}}-{{\varepsilon }_{cv}})\,
       ({{\varepsilon }_{nrw}}-{{\varepsilon }_{abv}})}\,+
\\ &&
\sum_{abcmnr} \frac{{z_{cr}}{\tilde{g}}_{abcn}{\tilde{g}}_{nrbm}
      {\tilde{g}}_{mwav}}{({{\varepsilon }_{mw}}-{{\varepsilon }_{av}})\,
      ({{\varepsilon }_{rw}}-{{\varepsilon }_{cv}})\,
      ({{\varepsilon }_{nrw}}-{{\varepsilon }_{abv}})}\,+
\\ &&
\sum_{abcmnr} \frac{{z_{cr}}{\tilde{g}}_{abcn}{\tilde{g}}_{nrmv}{g_{mwab}}}
    {({{\varepsilon }_{mw}}-{{\varepsilon }_{ab}})\,
      ({{\varepsilon }_{rw}}-{{\varepsilon }_{cv}})\,
      ({{\varepsilon }_{nrw}}-{{\varepsilon }_{abv}})}\,+
\\ &&
\sum_{abcmnr} \frac{{z_{cr}}{\tilde{g}}_{abcn}{\tilde{g}}_{rwbm}
      {\tilde{g}}_{mnav}}{({{\varepsilon }_{mn}}-{{\varepsilon }_{av}})\,
      ({{\varepsilon }_{rw}}-{{\varepsilon }_{cv}})\,
      ({{\varepsilon }_{nrw}}-{{\varepsilon }_{abv}})}\,+
\\ &&
\sum_{abcmnr} \frac{{z_{cr}}{\tilde{g}}_{abcn}{\tilde{g}}_{rwmv}{g_{mnab}}}
    {({{\varepsilon }_{mn}}-{{\varepsilon }_{ab}})\,
      ({{\varepsilon }_{rw}}-{{\varepsilon }_{cv}})\,
      ({{\varepsilon }_{nrw}}-{{\varepsilon }_{abv}})}\,
            + c.c.
\end{eqnarray*}

\begin{eqnarray*}
\lefteqn{Z_{1 \times 2}(D_{nl})=} \\
&&
\sum_{abcmnr}
\frac{{\tilde{g}}_{abmr}{z_{cn}}{\tilde{g}}_{nrcv}{g_{mwab}}}
    {({{\varepsilon }_{mr}}-{{\varepsilon }_{ab}})\,
      ({{\varepsilon }_{mw}}-{{\varepsilon }_{ab}})\,
      ({{\varepsilon }_{nr}}-{{\varepsilon }_{cv}})}\,+
\\ &&
-\sum_{abcmnr}
\frac{{\tilde{g}}_{abnr}{z_{cm}}{g_{nrcv}}{g_{mwab}}}
     {({{\varepsilon }_{mw}}-{{\varepsilon }_{ab}})\,
       ({{\varepsilon }_{nr}}-{{\varepsilon }_{ab}})\,
       ({{\varepsilon }_{nr}}-{{\varepsilon }_{cv}})}\,+
\\ &&
-\sum_{abcmnr}
\frac{{\tilde{g}}_{abnr}{z_{cm}}{\tilde{g}}_{rwcv}{g_{mnab}}}
     {({{\varepsilon }_{mn}}-{{\varepsilon }_{ab}})\,
       ({{\varepsilon }_{nr}}-{{\varepsilon }_{ab}})\,
       ({{\varepsilon }_{rw}}-{{\varepsilon }_{cv}})}\,+
\\ &&
\sum_{abcmnr}
\frac{{\tilde{g}}_{acmn}{z_{br}}{\tilde{g}}_{rwbc}{g_{mnav}}}
    {({{\varepsilon }_{mn}}-{{\varepsilon }_{ac}})\,
      ({{\varepsilon }_{mn}}-{{\varepsilon }_{av}})\,
      ({{\varepsilon }_{rw}}-{{\varepsilon }_{bc}})}\,+
\\ &&
-\sum_{abcmnr} \frac{{\tilde{g}}_{acmr}{z_{bn}}{\tilde{g}}_{nrbc}
       {\tilde{g}}_{mwav}}{({{\varepsilon }_{mr}}-{{\varepsilon }_{ac}})\,
       ({{\varepsilon }_{mw}}-{{\varepsilon }_{av}})\,
       ({{\varepsilon }_{nr}}-{{\varepsilon }_{bc}})}\,+
\\ &&
\sum_{abcmnr}
\frac{{\tilde{g}}_{acnr}{z_{bm}}{g_{nrbc}}{\tilde{g}}_{mwav}}
    {({{\varepsilon }_{mw}}-{{\varepsilon }_{av}})\,
      ({{\varepsilon }_{nr}}-{{\varepsilon }_{ac}})\,
      ({{\varepsilon }_{nr}}-{{\varepsilon }_{bc}})}\,+
\\ &&
\sum_{abcmnr} \frac{{\tilde{g}}_{acnr}{z_{bm}}{\tilde{g}}_{rwbc}
      {\tilde{g}}_{mnav}}{({{\varepsilon }_{mn}}-{{\varepsilon }_{av}})\,
      ({{\varepsilon }_{nr}}-{{\varepsilon }_{ac}})\,
      ({{\varepsilon }_{rw}}-{{\varepsilon }_{bc}})}\,  + c.c.
\end{eqnarray*}

\begin{eqnarray*}
\lefteqn{Z_{0 \times 3}(D_{nl})=} \\
&&
\sum_{abcmnr}
\frac{{z_{an}}{\tilde{g}}_{bcmr}{g_{rwbc}}{\tilde{g}}_{mnav}}
    {({{\varepsilon }_{mn}}-{{\varepsilon }_{av}})\,
      ({{\varepsilon }_{nw}}-{{\varepsilon }_{av}})\,
      ({{\varepsilon }_{rw}}-{{\varepsilon }_{bc}})}\,+
\\ &&
-\sum_{abcmnr}
\frac{{z_{ar}}{\tilde{g}}_{bcmn}{g_{nrbc}}{\tilde{g}}_{mwav}}
     {({{\varepsilon }_{mw}}-{{\varepsilon }_{av}})\,
       ({{\varepsilon }_{rw}}-{{\varepsilon }_{av}})\,
       ({{\varepsilon }_{nr}}-{{\varepsilon }_{bc}})}\,+
\\ &&
-\sum_{abcmnr}
\frac{{z_{ar}}{\tilde{g}}_{bcmn}{g_{rwbc}}{g_{mnav}}}
     {({{\varepsilon }_{mn}}-{{\varepsilon }_{av}})\,
       ({{\varepsilon }_{rw}}-{{\varepsilon }_{av}})\,
       ({{\varepsilon }_{rw}}-{{\varepsilon }_{bc}})}\,+
\\ &&
\sum_{abcmnr}
\frac{{z_{bm}}{\tilde{g}}_{acnr}{g_{nrcv}}{\tilde{g}}_{mwab}}
   {({{\varepsilon }_{mw}}-{{\varepsilon }_{ab}})\,
     ({{\varepsilon }_{mw}}-{{\varepsilon }_{bv}})\,
      ({{\varepsilon }_{nr}}-{{\varepsilon }_{cv}})}\,+
\\ &&
-\sum_{abcmnr} \frac{{z_{bn}}{\tilde{g}}_{acmr}{\tilde{g}}_{rwcv}
       {\tilde{g}}_{mnab}}{({{\varepsilon }_{mn}}-{{\varepsilon }_{ab}})\,
       ({{\varepsilon }_{nw}}-{{\varepsilon }_{bv}})\,
       ({{\varepsilon }_{rw}}-{{\varepsilon }_{cv}})}\,+
\\ &&
\sum_{abcmnr} \frac{{z_{br}}{\tilde{g}}_{acmn}{\tilde{g}}_{nrcv}
      {\tilde{g}}_{mwab}}{({{\varepsilon }_{mw}}-{{\varepsilon }_{ab}})\,
      ({{\varepsilon }_{rw}}-{{\varepsilon }_{bv}})\,
      ({{\varepsilon }_{nr}}-{{\varepsilon }_{cv}})}\,+
\\ &&
\sum_{abcmnr}
\frac{{z_{br}}{\tilde{g}}_{acmn}{\tilde{g}}_{rwcv}{g_{mnab}}}
    {({{\varepsilon }_{mn}}-{{\varepsilon }_{ab}})\,
      ({{\varepsilon }_{rw}}-{{\varepsilon }_{bv}})\,
      ({{\varepsilon }_{rw}}-{{\varepsilon }_{cv}})}\,  + c.c.
\end{eqnarray*}

Finally, the normalization correction due to valence triple excitations is defined as
\[
Z_\mathrm{norm}(T_v) =
- \frac{1}{2} \, \left( N_v^{(3)}(T_v) + N_w^{(3)}(T_v) \right) \, z_{wv} \,.
\]
The correction to normalization may be represented as
(the terms denoted $c.c.$ are to be calculated by taking the hermitian conjugate
of all preceding contributions)
\begin{eqnarray*}
\lefteqn{N_v^{(3)}(T_v) =} \\
&&
\sum_{abmnr} \frac{{\tilde{g}}_{abnr} \,{g_{nrmv}}\,{g_{mvab}}\,\,}
    {({{\varepsilon }_{mv}}-{{\varepsilon }_{ab}})\,
      {({{\varepsilon }_{nr}}-{{\varepsilon }_{ab}})}^2}\,+
\\ &&
\sum_{abcmn} \frac{{\tilde{g}}_{bcmn}\, {\tilde{g}}_{ancv}\,
      {\tilde{g}}_{mvab}\,\,}{{({{\varepsilon }_{mn}}-
         {{\varepsilon }_{bc}})}^2\,
      ({{\varepsilon }_{mv}}-{{\varepsilon }_{ab}})}\,+
\\ &&
\sum_{abmnr} \frac{{\tilde{g}}_{abnr}\,{g_{nrbm}}\,{\tilde{g}}_{mvav}\,\,}
    {({{\varepsilon }_m}-{{\varepsilon }_{a}})\,
      {({{\varepsilon }_{nr}}-{{\varepsilon }_{ab}})}^2}\,+
\\ &&
\sum_{abcmn} \frac{{\tilde{g}}_{bcmn}\,{g_{anbc}}\,{\tilde{g}}_{mvav}\,\,}
    {({{\varepsilon }_m}-{{\varepsilon }_{a}})\,
      {({{\varepsilon }_{mn}}-{{\varepsilon }_{bc}})}^2}\,+
\\ &&
\sum_{abmnr} \frac{{\tilde{g}}_{abnr}{\tilde{g}}_{rvbm}{\tilde{g}}_{mnav}}
    {({{\varepsilon }_{mn}}-{{\varepsilon }_{av}})\,
      {({{\varepsilon }_{nr}}-{{\varepsilon }_{ab}})}^2}\,+
\\ &&
\sum_{abmnr} \frac{{\tilde{g}}_{abnr}{\tilde{g}}_{rvmv}{g_{mnab}}}
    {({{\varepsilon }_{mn}}-{{\varepsilon }_{ab}})\,
      {({{\varepsilon }_{nr}}-{{\varepsilon }_{ab}})}^2}\,+
\\ &&
-\sum_{abcmn} \frac{{\tilde{g}}_{bcmn}{g_{avbc}}{g_{mnav}}}
     {({{\varepsilon }_{mn}}-{{\varepsilon }_{av}})\,
       {({{\varepsilon }_{mn}}-{{\varepsilon }_{bc}})}^2}\,+
\\ &&
-\sum_{abcmn} \frac{{\tilde{g}}_{bcmn}{\tilde{g}}_{avcv}{g_{mnab}}}
     {({{\varepsilon }_{mn}}-{{\varepsilon }_{ab}})\,
       {({{\varepsilon }_{mn}}-{{\varepsilon }_{bc}})}^2}\,
       + c.c.
\end{eqnarray*}

%\bibliographystyle{revtex}
%\bibliography{pnc,general,mypub,exact,vdW}

\begin{thebibliography}{18}
\expandafter\ifx\csname natexlab\endcsname\relax\def\natexlab#1{#1}\fi
\expandafter\ifx\csname bibnamefont\endcsname\relax
  \def\bibnamefont#1{#1}\fi
\expandafter\ifx\csname bibfnamefont\endcsname\relax
  \def\bibfnamefont#1{#1}\fi
\expandafter\ifx\csname citenamefont\endcsname\relax
  \def\citenamefont#1{#1}\fi
\expandafter\ifx\csname url\endcsname\relax
  \def\url#1{\texttt{#1}}\fi
\expandafter\ifx\csname urlprefix\endcsname\relax\def\urlprefix{URL }\fi
\providecommand{\bibinfo}[2]{#2}
\providecommand{\eprint}[2][]{\url{#2}}

\bibitem[{\citenamefont{Wood et~al.}(1997)\citenamefont{Wood, Bennett, Cho,
  Masterson, Roberts, Tanner, and Wieman}}]{WooBenCho97}
\bibinfo{author}{\bibfnamefont{C.~S.} \bibnamefont{Wood}},
  \bibinfo{author}{\bibfnamefont{S.~C.} \bibnamefont{Bennett}},
  \bibinfo{author}{\bibfnamefont{D.}~\bibnamefont{Cho}},
  \bibinfo{author}{\bibfnamefont{B.~P.} \bibnamefont{Masterson}},
  \bibinfo{author}{\bibfnamefont{J.~L.} \bibnamefont{Roberts}},
  \bibinfo{author}{\bibfnamefont{C.~E.} \bibnamefont{Tanner}},
  \bibnamefont{and} \bibinfo{author}{\bibfnamefont{C.~E.}
  \bibnamefont{Wieman}}, \bibinfo{journal}{Science}
  \textbf{\bibinfo{volume}{275}}, \bibinfo{pages}{1759} (\bibinfo{year}{1997}).

\bibitem[{\citenamefont{Bennett and Wieman}(1999)}]{BenWie99}
\bibinfo{author}{\bibfnamefont{S.~C.} \bibnamefont{Bennett}} \bibnamefont{and}
  \bibinfo{author}{\bibfnamefont{C.~E.} \bibnamefont{Wieman}},
  \bibinfo{journal}{Phys.\ Rev.\ Lett.} \textbf{\bibinfo{volume}{82}},
  \bibinfo{pages}{2484} (\bibinfo{year}{1999}).

\bibitem[{the()}]{theorPNCallNew}
\bibinfo{note}{A. Derevianko, Phys. Rev. Lett. {\bf 85}, 1618 (2000); V. A.
  Dzuba {\it et al.}, Phys. Rev. A {\bf 63}, 044103 (2001); M. G. Kozlov {\it
  et al.}, Phys. Rev. Lett. {\bf 86}, 3260 (2001); W. R. Johnson {\it et al.},
  Phys. Rev. Lett. {\bf 87}, 233001 (2001); A. I. Milstein and O. P. Sushkov,
  e-print: hep-ph/0109257; A. Derevianko, Phys. Rev A. {\bf 65}, 012106 (2002);
  Dzuba {\em et al.}, e-print: hep-ph/0111019}.

\bibitem[{\citenamefont{Dzuba et~al.}(1989)\citenamefont{Dzuba, Flambaum, and
  Sushkov}}]{DzuFlaSus89}
\bibinfo{author}{\bibfnamefont{V.~A.} \bibnamefont{Dzuba}},
  \bibinfo{author}{\bibfnamefont{V.~V.} \bibnamefont{Flambaum}},
  \bibnamefont{and} \bibinfo{author}{\bibfnamefont{O.~P.}
  \bibnamefont{Sushkov}}, \bibinfo{journal}{Phys.\ Lett.\ A}
  \textbf{\bibinfo{volume}{141}}, \bibinfo{pages}{147} (\bibinfo{year}{1989}).

\bibitem[{\citenamefont{Blundell et~al.}(1990)\citenamefont{Blundell, Johnson,
  and Sapirstein}}]{BluJohSap90}
\bibinfo{author}{\bibfnamefont{S.~A.} \bibnamefont{Blundell}},
  \bibinfo{author}{\bibfnamefont{W.~R.} \bibnamefont{Johnson}},
  \bibnamefont{and}
  \bibinfo{author}{\bibfnamefont{J.}~\bibnamefont{Sapirstein}},
  \bibinfo{journal}{Phys.\ Rev.\ Lett.} \textbf{\bibinfo{volume}{65}},
  \bibinfo{pages}{1411} (\bibinfo{year}{1990}), \bibinfo{note}{{P}hys.\ Rev.\ D
  {\bf 45}, 1602 (1992)}.

\bibitem[{\citenamefont{Amusia and Cherepkov}(1975)}]{AmuChe75}
\bibinfo{author}{\bibfnamefont{M.~Y.} \bibnamefont{Amusia}} \bibnamefont{and}
  \bibinfo{author}{\bibfnamefont{N.~A.} \bibnamefont{Cherepkov}},
  \bibinfo{journal}{Case Studies in Atomic Physics}
  \textbf{\bibinfo{volume}{5}}, \bibinfo{pages}{47} (\bibinfo{year}{1975}).

\bibitem[{\citenamefont{Sapirstein}(1998)}]{Sap98}
\bibinfo{author}{\bibfnamefont{J.}~\bibnamefont{Sapirstein}},
  \bibinfo{journal}{Rev.\ Mod.\ Phys.} \textbf{\bibinfo{volume}{70}},
  \bibinfo{pages}{55} (\bibinfo{year}{1998}).

\bibitem[{\citenamefont{Johnson et~al.}(1996)\citenamefont{Johnson, Liu, and
  Sapirstein}}]{JohLiuSap96}
\bibinfo{author}{\bibfnamefont{W.~R.} \bibnamefont{Johnson}},
  \bibinfo{author}{\bibfnamefont{Z.~W.} \bibnamefont{Liu}}, \bibnamefont{and}
  \bibinfo{author}{\bibfnamefont{J.}~\bibnamefont{Sapirstein}},
  \bibinfo{journal}{At.\ Data Nucl.\ Data Tables}
  \textbf{\bibinfo{volume}{64}}, \bibinfo{pages}{279} (\bibinfo{year}{1996}).

\bibitem[{\citenamefont{\v{C}\`{i}\v{z}ek}(1966)}]{Ciz66}
\bibinfo{author}{\bibfnamefont{J.}~\bibnamefont{\v{C}\`{i}\v{z}ek}},
  \bibinfo{journal}{J. Chem.\ Phys.} \textbf{\bibinfo{volume}{45}},
  \bibinfo{pages}{4256} (\bibinfo{year}{1966}).

\bibitem[{\citenamefont{Coester and K\"{u}mmel}(1960)}]{CoeKum60}
\bibinfo{author}{\bibfnamefont{F.}~\bibnamefont{Coester}} \bibnamefont{and}
  \bibinfo{author}{\bibfnamefont{H.~G.} \bibnamefont{K\"{u}mmel}},
  \bibinfo{journal}{Nucl.\ Phys.} \textbf{\bibinfo{volume}{17}},
  \bibinfo{pages}{477} (\bibinfo{year}{1960}).

\bibitem[{\citenamefont{Lindgren and Morrison}(1986)}]{LinMor86}
\bibinfo{author}{\bibfnamefont{I.}~\bibnamefont{Lindgren}} \bibnamefont{and}
  \bibinfo{author}{\bibfnamefont{J.}~\bibnamefont{Morrison}},
  \emph{\bibinfo{title}{Atomic Many--Body Theory}}
  (\bibinfo{publisher}{Springer--Verlag}, \bibinfo{address}{Berlin},
  \bibinfo{year}{1986}), \bibinfo{edition}{2nd} ed.

\bibitem[{\citenamefont{Blundell et~al.}(1989)\citenamefont{Blundell, Johnson,
  Liu, and Sapirstein}}]{BluJohLiu89}
\bibinfo{author}{\bibfnamefont{S.~A.} \bibnamefont{Blundell}},
  \bibinfo{author}{\bibfnamefont{W.~R.} \bibnamefont{Johnson}},
  \bibinfo{author}{\bibfnamefont{Z.~W.} \bibnamefont{Liu}}, \bibnamefont{and}
  \bibinfo{author}{\bibfnamefont{J.}~\bibnamefont{Sapirstein}},
  \bibinfo{journal}{Phys.\ Rev.\ A} \textbf{\bibinfo{volume}{40}},
  \bibinfo{pages}{2233} (\bibinfo{year}{1989}).

\bibitem[{\citenamefont{Blundell et~al.}(1991)\citenamefont{Blundell, Johnson,
  and Sapirstein}}]{BluJohSap91}
\bibinfo{author}{\bibfnamefont{S.~A.} \bibnamefont{Blundell}},
  \bibinfo{author}{\bibfnamefont{W.~R.} \bibnamefont{Johnson}},
  \bibnamefont{and}
  \bibinfo{author}{\bibfnamefont{J.}~\bibnamefont{Sapirstein}},
  \bibinfo{journal}{Phys.\ Rev.\ A} \textbf{\bibinfo{volume}{43}},
  \bibinfo{pages}{3407} (\bibinfo{year}{1991}).

\bibitem[{\citenamefont{Safronova et~al.}(1999)\citenamefont{Safronova,
  Johnson, and Derevianko}}]{SafJohDer99}
\bibinfo{author}{\bibfnamefont{M.~S.} \bibnamefont{Safronova}},
  \bibinfo{author}{\bibfnamefont{W.~R.} \bibnamefont{Johnson}},
  \bibnamefont{and}
  \bibinfo{author}{\bibfnamefont{A.}~\bibnamefont{Derevianko}},
  \bibinfo{journal}{Phys.\ Rev.\ A} \textbf{\bibinfo{volume}{60}},
  \bibinfo{pages}{4476} (\bibinfo{year}{1999}).

\bibitem[{\citenamefont{Blundell et~al.}(1987)\citenamefont{Blundell, Guo,
  Johnson, and Sapirstein}}]{BluGuoJoh87}
\bibinfo{author}{\bibfnamefont{S.~A.} \bibnamefont{Blundell}},
  \bibinfo{author}{\bibfnamefont{D.~S.} \bibnamefont{Guo}},
  \bibinfo{author}{\bibfnamefont{W.~R.} \bibnamefont{Johnson}},
  \bibnamefont{and}
  \bibinfo{author}{\bibfnamefont{J.}~\bibnamefont{Sapirstein}},
  \bibinfo{journal}{At.\ Data Nucl.\ Data Tables}
  \textbf{\bibinfo{volume}{37}}, \bibinfo{pages}{103} (\bibinfo{year}{1987}).

\bibitem[{\citenamefont{Wolfram}(1999)}]{Wol99}
\bibinfo{author}{\bibfnamefont{S.}~\bibnamefont{Wolfram}},
  \emph{\bibinfo{title}{The Mathematica Book}} (\bibinfo{publisher}{Wolfram
  Media/Cambridge University Press, Champaign, Illinois},
  \bibinfo{year}{1999}), \bibinfo{edition}{4th} ed.

\bibitem[{\citenamefont{Bishop and K\"{u}mmel}(1987)}]{BisKum87}
\bibinfo{author}{\bibfnamefont{R.~F.} \bibnamefont{Bishop}} \bibnamefont{and}
  \bibinfo{author}{\bibfnamefont{H.~G.} \bibnamefont{K\"{u}mmel}},
  \bibinfo{journal}{Physics Today} \textbf{\bibinfo{volume}{3}},
  \bibinfo{pages}{52} (\bibinfo{year}{1987}).

\bibitem[{\citenamefont{Safronova et~al.}(1998)\citenamefont{Safronova,
  Derevianko, and Johnson}}]{SafDerJoh98}
\bibinfo{author}{\bibfnamefont{M.~S.} \bibnamefont{Safronova}},
  \bibinfo{author}{\bibfnamefont{A.}~\bibnamefont{Derevianko}},
  \bibnamefont{and} \bibinfo{author}{\bibfnamefont{W.~R.}
  \bibnamefont{Johnson}}, \bibinfo{journal}{Phys.\ Rev.\ A}
  \textbf{\bibinfo{volume}{58}}, \bibinfo{pages}{1016} (\bibinfo{year}{1998}).

\end{thebibliography}

\end{document}